\documentclass[12pt,preprint]{aastex}
\usepackage{graphicx}
\shorttitle{The three parameters of the HB morphology in GCs} 
\shortauthors{A.\ P.\ Milone, et al.\ } 
\usepackage{ulem}
\usepackage{txfonts}
\usepackage{longtable,lscape}

\begin{document}
\title{Global and non-global parameters of horizontal branch morphology of globular clusters
          \footnote{           Based on observations with  the
                               NASA/ESA {\it Hubble Space Telescope},
                               obtained at  the Space Telescope Science
                               Institute,  which is operated by AURA, Inc.,
                               under NASA contract NAS 5-26555.}}
\author{
 A.\, P.\, Milone\altaffilmark{2,3,4},
 A.\, F.\, Marino\altaffilmark{2},
 A.\, Dotter\altaffilmark{2},
 J.\, E.\, Norris\altaffilmark{2}, 
 H.\, Jerjen\altaffilmark{2}, 
 G.\, Piotto\altaffilmark{5,7}, 
 S.\, Cassisi\altaffilmark{6}, 
 L.\, R.\, Bedin\altaffilmark{7}, 
 A.\, Recio Blanco\altaffilmark{8},
 A.\, Sarajedini\altaffilmark{9}, 
 M.\, Asplund\altaffilmark{2},
 M.\, Monelli\altaffilmark{3,4}, 
 A.\, Aparicio\altaffilmark{3,4} 
 }

\altaffiltext{2}{Research School of Astronomy and Astrophysics, The Australian National University, Cotter Road, Weston, ACT, 2611, Australia; milone@mso.anu.edu.au, amarino@mso.anu.edu.au, jen@mso.anu.edu.au, jerjen@mso.anu.edu.au, martin@mso.anu.edu.au}

\altaffiltext{3}{Instituto de Astrof\`\i sica de Canarias, E-38200 La Laguna, Tenerife, Canary Islands, Spain; monelli@iac.es, aparicio@iac.es}

\altaffiltext{4}{Department of Astrophysics, University of La Laguna, E-38200 La Laguna, Tenerife, Canary Islands, Spain}

\altaffiltext{5}{Dipartimento  di  Fisica e Astronomia, `Galileo Galilei' Universit\`a  di Padova, Vicolo dell'Osservatorio 3, Padova I-35122, Italy; giampaolo.piotto@unipd.it }

\altaffiltext{6}{INAF-Osservatorio Astronomico di Collurania, via Mentore Maggini, I-64100 Teramo, Italy; cassisi@oa-teramo.inaf.it} 

\altaffiltext{7}{INAF-Osservatorio Astronomico di Padova, Vicolo dell'Osservatorio 5, I-35122 Padua, Italy; luigi.bedin@oapd.inaf.it}

\altaffiltext{8}{Observatoire de la Cote d'Azur, UMR Cassiopee, B.\ P.\ 4229, 06304 Nice Cedex 04, France; alejandra.recio-blanco@oca.eu}

\altaffiltext{9}{Department of Astronomy, University of Florida, 211 Bryant Space Science Center, Gainesville, FL 32611, USA; ata@astro.ufl.edu}

\shorttitle{HB morphology GCs} 
\shortauthors{Milone et al.\ } 

\begin{abstract}
\noindent The horizontal branch (HB) morphology of globular clusters (GCs) is mainly determined by metallicity. However, the fact that GCs with almost the same metallicity exhibit different HB morphologies demonstrates that at least one more parameter is needed to explain the HB morphology. It has been suggested that one of these should be a {\it global} parameter that varies from GC to GC, and the other a {\it non-global} parameter that varies within the GC.\\
 In this study we provide empirical evidence corroborating this idea. 
  We used the photometric catalogs obtained with the 
 Advanced Camera for Surveys (ACS) of the {\it Hubble Space Telescope} ({\it HST}) and analyse the CMDs of 74 GCs.
 The HB morphology of our sample of GCs has been investigated on the basis of the two new parameters $L1$ and $L2$ that measure the distance between the RGB and the coolest part of the HB, and the color extension of the HB, respectively.\\
We find that $L1$ correlates with both metallicity and age, whereas $L2$ most strongly correlates with the mass of the hosting GC.
The range of helium abundance among the stars in a GC, characterised by $\Delta$Y and associated with the presence of multiple stellar populations, has been estimated in a few GCs to date. In these GCs we find a close relationship among $\Delta$Y, GC mass, and $L2$. We conclude that age and metallicity are the main {\it global} parameters 
while the range of helium abundance within a GC is the main {\it non-global} parameter defining the HB morphology of Galactic GCs.
\end{abstract}

\keywords{stars: abundances ---  stars: Population II --- globular clusters --- general}

\section{Introduction}
\label{introduction}
Since the early 1950s, metallicity has been considered the main parameter 
 that determines the horizontal-branch (HB) morphology in globular clusters
 (GCs; e.g., Arp\,1952).
Within a few years, evidence that some GCs with similar metallicity exhibit different HB morphologies suggested that at least a second parameter (2ndP) is required to properly characterise the HB morphology of GCs (e.g.\,Sandage \& Wallerstein\, 1960; van den Bergh\, 1965). 
Since then, the so called 2ndP problem has been widely investigated by many authors. 

Several candidates have been suggested as possible 2ndPs, 
including mass loss (e.g.\,Peterson\,1982; Catelan\,2000), stellar rotation (e.g.\,Mengel \& Gross\,1976; Fusi-Pecci \& Renzini\,1978), 
planetary systems (e.g.\,Soker\,1998; Siess \& Livio\,1999),
magnetic fields (e.g.\,Rood \& Seitzer\,1981), 
and GC ellipticity (Norris\, 1983) 
but a comprehensive picture is still lacking. 
Age  (e.g.\,Searle \& Zinn\,1978; Catelan \& de Fretais Pacheco\, 1993; Lee et al.\,1994), GC central density  (e.g.\,Fusi-Pecci et al.\,1993), 
GC mass  (e.g.\,Recio-Blanco et al.\,2006), and helium abundance 
 (e.g.\,Sandage \& Wildey\,1967; van den Bergh\,1967) are among the best candidates. We refer the reader to the papers by  Freeman \& Norris\,(1981), Catelan et al.\,(2009), Dotter et al.\,(2010), Gratton et al.\,(2010), and references therein for reviews on HB stars and the 2ndP phenomenon in GCs.

The study of GCs has changed dramatically in recent years due to the overwhelming evidence for the existence of multiple stellar populations in GCs. In this way of thinking a GC is made up of a first generation of stars, formed from the GC's primordial gas cloud, and at least one later generation, formed from a dilution of the primordial gas and the chemical yields of the high- and intermediate-mass stars of the first generation.

The possibility of GC self-enrichment, especially as it relates to enhanced helium, as cause for the variation of the HB morphology has been investigated by several authors, as multiple stellar populations with different helium abundance can indeed explain features such as tails and multimodalities in the HBs of GCs  
 (e.g.\,Ferraro et al.\,1998; D'Antona et al.\,2002, 2005; Piotto et al.\,2007;  Caloi \& D'Antona\,2008; Gratton et al.\,2010).
The idea of a connection between multiple stellar populations and HB morphology arose in the early 1980s, when pioneering papers showed that the cyanogen distribution is closely connected to the shape of the HB 
 (e.g.\,Norris\,1981; Norris et al.\,1981; Smith \& Norris\, 1993)
this result has been confirmed by recent studies of HB stars.

In this context the GC M\,4 is exemplary. 
High-resolution spectroscopy of red-giant branch (RGB) stars reveals that this GC hosts two stellar populations with different Na and O abundances while photometry reveals two RGBs in the $U$ versus $U-B$ color-magnitude diagram (CMD). 
The HB of M\,4 is bimodal and well-populated on both sides of the RR Lyrae gap 
 (Marino et al.\,2008).
The bimodality in Na and O is also present among the HB stars. Blue-HB stars belong to the second population and are O-poor and Na-rich, while red HB stars are first population (Marino et al.\,2008, 2011a).
 Similar analysis of Na and O in HB stars in other GCs show that first generation HB stars preferentially populate the reddest HB segment while second generation HB stars tend towards bluer colors 
 (Villanova et al.\,2009, Gratton et al.\,2011, 2012, 2013), Lovisi et al.\,2012, Marino et al.\,2013a).
More recently, 
 Marino et al.\,(2013b)
 inferred from direct spectroscopic measurements that Na-rich HB stars of NGC\,2808 are also strongly helium enhanced.

Several factors influence HB morphology and it is difficult to disentangle the different effects.
An important point in the study of the 2ndP is that the metric used to characterise HB morphology is not objective: the chosen way of representing the HB stars in a GC as a number has a non-trivial influence on the results of the investigation. Most studies to date adopt a single HB morphology metric. Different studies, using different metrics, can easily reach conflicting conclusions about the identity of the 2ndP.

Consider the following two examples. 
 Recio-Blanco et al.\,(2006) 
 defined their HB morphology metric as the maximum effective temperature along the HB and found that more massive GCs tend to have hotter HBs. 
 Dotter et al.\,(2010) 
measured the median color difference between the HB and the RGB at the level of the HB ($\Delta$($V-I$)) 
from {\it Hubble Space Telescope} ({\it HST}) Advanced Camera for Surveys (ACS) photometry of sixty GCs, and demonstrated that, after the metallicity dependence is accurately removed, $\Delta$($V-I$) correlates most strongly with GC age.

 Dotter et al.\,(2010, see their Figure~2) 
compared $\Delta$($V-I$) with the widely-used HB Type index\footnote{Defined as (B$-$R)/(B+V+R), where B, R, and V are the numbers of blue HB, red HB, and variables stars. It measures the relative contributions of stars bluer or redder than the RR Lyrae instability strip, and is variously known by the definition, HB Type, or HB Ratio (HBR).} and the maximum effective temperature along the HB as defined by 
 Recio-Blanco et al.\,(2006).
The comparison shows that $\Delta$($V-I$) and HB Type are closely correlated but that $\Delta$($V-I$) and Recio-Blanco et al.'s maximum HB temperature are not: they have very different sensitivities.
It is not surprising that a metric sensitive to the extremes of the distribution  correlates with a different (second) parameter than a metric sensitive to the center of the distribution. The problem is that 2ndP studies typically select only one HB morphology metric and the conclusions are influenced by that choice.

We suggest that a more effective way to proceed is to consider more than one HB morphology metric simultaneously. For maximum effect, these metrics should share a simple, common definition but not be closely correlated with each other. The motivation for this approach is not only based on the practicalities outlined above.
 Freeman \& Norris\,(1981) 
argued that two parameters, one global and one local, may be needed to fully describe the observed variations in HB morphology.  The local parameter is one that varies within a single GC; the global parameter is one that varies among the GC population.

 The aim of this paper is an empirical investigation of the parameters governing the HB morphology of GCs, in the context of the classical 2ndP phenomenon.
 To do this we use the homogeneous high-accuracy photometry  from GO\,10775, the
 Advanced Camera for Surveys (ACS) Survey of Galactic GCs 
 (P.I.\,A.\,Sarajedini, Sarajedini et al.\,2007),
and from GO\,11586 
 (P.I.\,A.\,Dotter, Dotter et al.\,2011)
and additional photometry from {\it HST} to re-investigate the HB morphology in GCs in light  of the new findings on multiple stellar populations in GCs and of the  global versus non-global parameter idea by 
 Freeman \& Norris\,(1981).
The paper is organised as follows. Sect.~\ref{sect:data} describes the observational data. Sect.~\ref{sect:morphology} introduces the quantities adopted to describe the HB morphology and defines the new HB morphology parameters $L1$ and $L2$. Sect.~\ref{GCpar} assembles a variety of GC parameters from the literature. Sect.~\ref{results} compares these parameters with $L1$ and $L2$. Sect.~\ref{literature} discusses our findings in the context of similar studies in the literature. Finally, we summarise our findings in Sect.~\ref{summary}.

\section{The data sample}
\label{sect:data}
We used the photometric catalogs obtained from GO\,10775 and GO\,11586 that include homogeneous photometric and astrometric measurements for 65 and 6 GCs, respectively.  For each of them, the dataset consists of one short and four or five long  exposures in the F606W and F814W bandpasses. We excluded from GO\,10775 three GCs: Pal\,1 and E\,3 for the lack of identifiable HB stars, and Pal\,2 for the extreme differential reddening.
The details concerning the data, the data reduction, and the calibration are given in Anderson et al.\,(2008) and Dotter et al.\,(2011).

In their study of the HB in GCs, 
 Dotter et al.\,(2010) emphasised the importance of properly accounting for the outer Halo, where the 2ndP is more evident. 
 To increase the number of outer-Halo GCs, we have extended the GO\,10775 and GO\,11586 to six more GCs: AM-1, Eridanus, NGC\,2419, Pal\,3, Pal\,4, Pal\,14.
For both Pal\,4 and Pal\,14 we used 2$\times$60s F606W and 2$\times$80s F814W  ACS/WFC images from GO\,10622 (PI.\, Dolphin), while for NGC\,2419 we used the F606W and F814W magnitudes published by 
 di Criscienzo et al.\,(2011). 
These data have been reduced as already described in  
 Anderson et al.\,(2008), Dotter et al.\,(2011), and di Criscienzo et al.\,(2011). 
 For Pal\,3 and Eridanus we used  ground-based $V$, $I$ photometry from 
 
 Stetson et al.\,(1999), 
and for AM-1 Wide Field Planetary Camera 2 (WFPC2) photometry from  
 Dotter et al.\,(2008)
 in F555W and F814W. Photometry for these three GCs has been transformed into F606W and F814W ACS/WFC bands by using the relationships given in 
 Sirianni et al.\,(2005).

Photometry has been corrected for spatial photometric zero-point variation both due to differential reddening and small inaccuracies in the PSF model 
 (Anderson et al.\,2008).
For most GCs we used the corrected magnitudes and colors published by 
 Milone et al.\,(2012b, 59 GCs), Piotto et al.\,(2012, NGC\,6752), di Criscienzo et al.\,(2011, NGC\,2419), and Bellini et al.\,(2010, $\omega$ Centauri). 
For the remaining GCs we corrected the photometry following the procedure described in Milone et al.\,(2012b).

\section{Two new parameters to describe the HB morphology: $L1$ and $L2$}
\label{sect:morphology}
To investigate the HB morphology, we defined two quantities: $L1$, the color difference between the RGB and the coolest border the HB, and $L2$, the color extension of the HB\footnote{We emphasise that we have not introduced $L1$ and $L2$ because we believe that they are more (or less) efficient than any other diagnostic previously used in the study of the HB morphology. Nevertheless, results of the empirical analysis presented in this paper will show that $L1$ and $L2$ are useful tools to shed light on the 2ndP phenomenon.
Any search for the best diagnostic of the HB morphology is obviously outside the scope of our paper and is possibly naive: one would need to understand a-priori which parameters determine HB morphology, and then compare the sensitivity of $L1$, $L2$, and the other diagnostics of the HB morphology used in literature with these parameters.}.

\begin{figure*}[ht!]
   \centering
   \epsscale{.75}
      \plotone{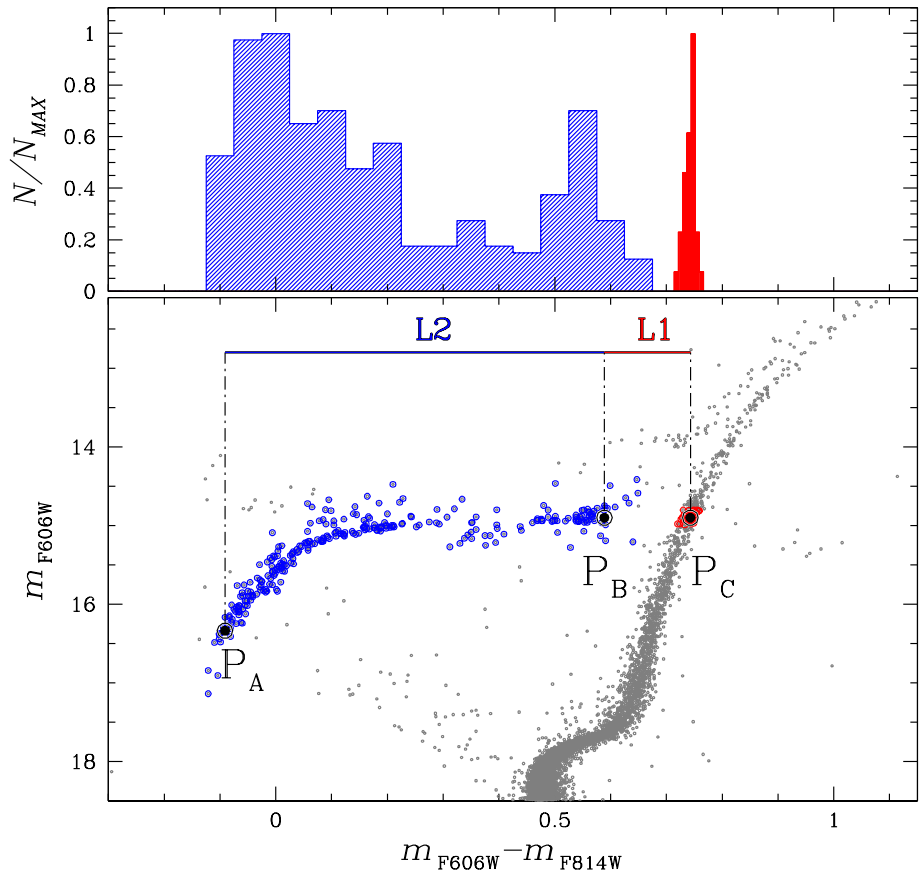}
      \caption{\textit{Upper panel:} Normalised histogram color distribution of
       stars in the HB (blue histogram) and RGB sample (red histogram) 
       for NGC\,5904.
       The RGB sample includes all the RGB stars with luminosity differing by less than $\pm$0.1 F606W mag from the mean level of the HB.
       The two samples of HB and RGB stars are colored blue and red, 
       respectively in the lower-panel CMD, where we also show 
       the points $P_{\rm A}$, $P_{\rm B}$, $P_{\rm C}$ 
       and the $L1$ and $L2$ segments (see text for details).}
         \label{procedure}
   \end{figure*}

The procedure to determine $L1$ and $L2$ is illustrated in Fig.~\ref{procedure} for the case of NGC\,5904 (M\,5).
We selected by eye a sample of HB stars that we plotted as blue 
circles in the lower panel, and a sample of RGB stars that we represented
with red points. The RGB sample includes all the RGB stars with luminosity
 differing by less than $\pm$0.1 F606W mag from the mean level of the HB 
($F606W_{\rm HB}$), where the $F606W_{\rm HB}$ values are taken from 
 Dotter et al.\,(2010, Table~1).
The histograms of the normalised $m_{\rm F606W}-m_{\rm F814W}$ color 
 distribution for the HB and RGB sample are shown in the upper panel and 
 colored blue and red, respectively.
 We have defined two points on the HB, $P_{\rm A}$ and $P_{\rm B}$, whose colors correspond to the fourth and the ninety-sixth percentile of the color distribution of HB stars. The color of the third point $P_{\rm C}$ is assumed as the median color of RGB stars. $L1$ is defined as the color difference between $P_{\rm C}$ ans $P_{\rm B}$, and $L2$ as the color difference between $P_{\rm B}$ and $P_{\rm A}$.
Uncertainties on $P_{\rm A}$, $P_{\rm B}$,  $P_{\rm C}$, $L1$ and $L2$ are estimated for each GC 
by bootstrapping with replacements performed 1,000 times on both the RGB and the HB. The error bars indicate one standard deviation (68.27$^{\rm th}$ percentile) of the bootstrapped measurements.
The colors of $P_{\rm A}$, $P_{\rm B}$,  $P_{\rm C}$, the values of $L1$ and $L2$,
 and the corresponding errors,  are listed in Tab.~1.

 RR\,Lyrae have been observed at random phases and some of them could lie outside the instability strip.
Similarly to what has been done in previous papers on the second parameter (e.g.\, Gratton et al.\, 2010, Dotter et al.\,2010) we included in our analysis only those RR\,Lyrae that are close to the instability strip. In the appendix  we investigate the impact of excluding RR\,Lyrae that might be out of the instability strip in the determination of $L1$ and $L2$ and conclude that this does not affect the conclusions of our paper.

\section{GC parameters}
\label{GCpar}
In the next section we  shall compare the $L1$ and $L2$ HB morphology indicators with the physical and morphological GC parameters described here.
Metallicity ([Fe/H]), absolute visual magnitude ($M_{\rm V}$), central velocity dispersion ($\sigma_{\rm V}$), ellipticity ($\epsilon$), central concentration ($c$), core relaxation timescale ($\tau_{\rm c}$), half-mass relaxation timescale ($\tau_{\rm hm}$), logarithm of central
 stellar density ($\rho_{0}$), central surface brightness ($\mu_{\rm V}$), and Galactocentric distance ($R_{\rm GC}$) are extracted from the 2010 edition of the Harris\,(1996) catalog. The specific frequency of RR\,Lyrae variables ($S_{\rm RR\,Lyrae}$) is taken from the 2003 edition of the Harris\,(1996) catalog. The fraction of binary stars have been measured by Milone et al.\,(2012b,c) in the core of the GC ($f_{\rm bin}^{\rm C}$), in the region between the core and the half-mass radius ($f_{\rm bin}^{\rm C-HM}$), and outside the half-mass radius ($f_{\rm bin}^{\rm oHM}$).
We also use age and helium measurements, and some indicators of light-element variations, as discussed in detail in Sect.~\ref{age} and  Sect.~\ref{elio}, respectively.

\subsection{Age} 
\label{age}
We adopt GC age measurements from De Angeli et al.\,(2005), Mar{\'{\i}}n-Franch et al.\,(2009), VandenBerg et al.\,(2013), and Leaman et al.\,(2013), and Dotter et al.\,(2010, 2011).
 De Angeli et al.\,(2005) determined ages for 55 GCs by measuring the difference between the HB and the turnoff in two internally photometrically homogeneous {\it HST} and ground-based databases  (Piotto et al.\,2002; Rosemberg et al.\,1999).
 Among the GCs studied by  De Angeli et al.\,(2005),
41 are in common with the ACS data.

 Mar{\'{\i}}n-Franch et al.\,(2009) used the same GO\,10775 photometric database analysed in  the present paper to estimate relative ages for 64 GCs that are included in our dataset with internal uncertainties of 2\%-7\%. Ages are derived by comparing the observed relative position of the GC main sequence turnoff (MSTO) with an isochrones-based grid of MSTOs. 

 More recently, VandenBerg et al.\,(2013) and Leaman et al.\,(2013) compared Victoria-Regina isochrones with photometry from Sarajedini et al.\,(2007) and Anderson et al.\,(2008) to derive ages for 61 GCs that are also included in our paper. To do this, they adopted an improved version of the classical `vertical method', which is based on the luminosity difference between the MSTO and the zero-age HB.

Age estimates by Dotter et al.\,(2010, 2011)
 are determined by using isochrone fitting to the CMDs for 59 of the GCs studied in this paper; note that these authors omitted 6 GCs present in the ACS Survey of Galactic GCs due to the presence of multiple stellar populations  with either large helium variation or double subgiant branch (SGB).
 For the remainder of the ACS Survey GCs (NGC\,1851, NGC\,2808, NGC\,6388, NGC\,6441, NGC\,6656, and NGC\,6715) we calculated ages following the same recipes from Dotter and collaborators. We obtained: 11.00$\pm$0.50 Gyr (NGC\,1851), 11.50$\pm$0.75 Gyr (NGC\,2808), 11.75$\pm$0.75 Gyr (NGC\,6388), 12.00$\pm$1.00 Gyr (NGC\,6441), 13.50$\pm$1.00 Gyr (NGC\,6656), and 13.25$\pm$0.75 Gyr (NGC\,6715).

 NGC\,1851, NGC\,6388, NGC\,6656, and NGC\,6715 each exhibit a double SGB (Milone et al.\,2008; Marino et al.\,2009; Piotto et al.\,2012) that is consistent with two groups of stars that have either an age difference of 1-2 Gyrs or almost the same age and different C+N+O abundance (Cassisi et al.\,2008; Ventura et al.\,2009).
 High-resolution spectroscopy of SGB and RGB stars in NGC\,6656 showed that the faint SGB of this GC is indeed made of C+N+O enhanced stars (Marino et al.\,2011a; Alves-Brito et al.\,2012) and that, by accounting for the chemical content of the two stellar groups, isochrone fitting of the double SGB indicates that their ages do not differ by more than $\sim$300 Myr (Marino et al.\,2012).
Large star-to-star C+N+O variations, with faint SGB-stars being also enhanced in C+N+O, have been also observed in NGC\,1851 (Yong et al.\,2009 and in preparation) even if this result is not confirmed by  Villanova et al.\,(2010). 
For GCs with a double SGB we assume the age obtained from fitting the bright SGB. To distinguish between GCs with single or bimodal SGBs, the statistical analysis of the relation between HB-morphology and age presented in Sect.~\ref{results} will be presented with and without these double-SGB GCs.

  When investigating the effect of age on HB morphology a challenge  comes from the fact that the population of Milky Way GCs mainly consists of old objects. Young clusters are hence important in the present investigation as any effect of age on the HB morphology would be better identified when comparing clusters with large age differences.
In order to better sample the Galactic GC population at all Galactocentric distances,  Dotter et al.\,(2010, 2011) expanded the sample studied by  Mar{\'{\i}}n-Franch et al.\,(2009) to include several more-distant GCs: AM-1 and Pal\,14 (Dotter et al.\,2008); Pal\,3, Pal\,4, and Eridanus (Stetson et al.\,1999), and IC\,4499, Pal\,5, Pyxis, and Ruprecht\,106 (Dotter et al.\,2011). It come out that several of them are also young clusters.

  Ages from Dotter and collaborators are available for 73 GCs, but only 61, 41, and 61 of them are included in the age compilations published by 
Mar{\'{\i}}n-Franch  et al.\,(2009), De Angeli et al.\,(2005), and by VandenBerg et al.\,(2013) and Leaman et al.\,(2013), respectively. 
 Since the sample by Dotter and collaborators is the most complete and is based on the best dataset available, we began by using their ages to investigate possible relations with $L1$ and $L2$.
Although the other age compilations include a smaller GC sample, they can be used to demonstrate that our conclusions do not rely on a particular set of ages.

 Another challenge comes from age errors that can be as large as 1.25 Gyr for the Dotter et al.\,(2010, 2011) sample. To minimize the impact of error measurements, our conclusions are based on the average properties of groups of GCs that will be defined in Sect.~\ref{sub:FeH}.

\subsection{Light-element and helium variations}
\label{elio}
Our analysis makes use of several indicators of the  intracluster light-element variations as well as of measurements of the helium differences among the multiple stellar populations in GCs. Some of these parameters have been defined several decades ago and widely discussed in the context of the 2ndP phenomenon while others have been introduced more recently. 

 Norris\,(1987) defined the ratio ($R_{\rm CN}$) between CN-strong and CN-weak stars and provided $R_{\rm CN}$ measurements for a sample of 12 GCs, which was increased to 16 by  Smith \& Mateo\,(1990).
The interquartile range of the [O/Na] ratio, IQR[O/Na], can be considered as another indicator of the internal light-element variation to quantify the extension of the sodium-oxygen anticorrelation (Carretta et al.\,2006). This parameter is available for 24 GCs: for 14 GCs, namely NGC\,104, NGC\,1904, NGC\,2808, NGC\,3201, NGC\,4590, NGC\,5904, NGC\,6171, NGC\,6218, NGC\,6254, NGC\,6388, NGC\, 6441, NGC\,6809, NGC\,6838, and NGC\,7078, we used the IQR[O/Na] values listed by 
 Carretta et al.\,(2010a). 
For the other 10 GCs, we have calculated IQR[O/Na] by using oxygen and sodium abundances available in the literature.
We obtained IQR[O/Na]=0.61 and 0.67 for NGC\,288 and  NGC\,362, respectively (Shetrone \& Keane\,2000);
IQR[O/Na]=0.68 for NGC\,1851 (Villanova et al.\,2010); IQR[O/Na]=0.54 and 1.04 for NGC\,5272 and NGC\,6205, respectively (Sneden et al.\,2004);
IQR[O/Na]=0.28 for NGC\,6397 (Lind et al.\,2011); IQR[O/Na]=1.16 for NGC\,6715 (Carretta et al.\,2010a); IQR[O/Na]=0.91 for NGC\,6752  (Yong et al.\,2008); IQR[O/Na]=0.49, 0.67, and 1.07 for NGC\,6121, NGC\,6656, and NGC\,5139, respectively (Marino et al.\,2008,2011b,a).

In their survey of multiple stellar populations in GCs, Monelli et al.\,(2013) 
defined the photometric index $c_{\rm UBI}$=($U-B$)$-$($B-I$) and found that all the GCs they analysed show a multimodal or spread RGB in the $V$ versus $c_{\rm UBI}$ diagram, with the $c_{\rm UBI}$ value of each star being related to its light element abundances. The  $c_{\rm UBI}$ index width of the RGB ($W_{\rm RGB}$)  is listed by  Monelli et al.\,(2013) for 22 GCs. 

Some stars in GCs have light-element abundance similar to halo-field stars of the same metallicity (e.g.\,Kraft\,1994). It is widely accepted that these stars constitute the first stellar population and can be distinguished from the other GC stars either on the basis of their abundance of C, N, Na, O or their position in the CMD. 
 For 47\,Tuc, NGC\,6397, NGC\,6752, NGC\,288 we adopted  the fraction of first-population stars with respect to the total population ($f_{\rm POPI}$) determined from photometric studies (Milone et al.\,2012c,d, 2013; Piotto et al.\,2013).
For NGC\,2808, and NGC\,6121 we used the fractions derived by  Marino et al.\,(2008) and Marino et al.\,(2013b), and for NGC\,3201, NGC\,4590, NGC\,5272, NGC\,5904, NGC\,6171, NGC\,6205, NGC\,6218, NGC\,6254, NGC\,6388, NGC\,6496, NGC\,6838, NGC\,7078, and NGC\,7079 we used the value given by Carretta et al.\,(2010b). 

While the quantities $R_{\rm CN}$, IQR[O/Na], and $W_{\rm RGB}$ are mainly related to the internal variations of light-elements, recent analysis based on multi-wavelength photometry made it possible to estimate the helium difference between stellar populations in a single GC.
These studies have revealed that the CMD of a GC is typically composed of intertwined sequences, the separate identities of which can be followed continuously from the MS up to the RGB. These sequences are associated with stellar populations with different light-element and helium abundances 
 (e.g.\,Milone et al.\,2012d,a). The comparison of the observed multi-color difference between the different MSs and RGBs with colors obtained from appropriate theoretical stellar atmospheres provides an estimate of the maximum helium difference ($\Delta$Y) between the stellar populations that is not based on HB stars. Although this technique has been applied to only a few GCs to date, we shall use the available $\Delta$Y measurements, summarised in Tab.~2, to investigate possible relations with HB morphology in the next section.

While $\Delta Y$ indicates the maximum internal variation in helium, the R-parameter ($R$) defined by Iben\,(1968) as the number ratio of HB to RGB stars brighter than the HB level is sensitive to the initial helium content of GCs.  
At a given metallicity, a higher initial He-content implies a brighter HB and, in turn, a lower number of RGB stars. In this paper we used the values of $R$ determined by Salaris et al.\,(2004). Gratton et al.\,(2010) introduced a similar parameter R'=$N_{\rm HB}/N'_{\rm RGB}$, where $N_{\rm HB}$ is the number of HB stars, $N'_{\rm RGB}$ the number of RGB stars brighter than $V$(HB)+1, and $V$(HB) is the $V$ magnitude of the HB taken from the Harris\,(1996) catalog. Gratton and collaborators used the R-parameter method Iben\,(1968) to derive He abundance for GCs (Y(R')). As suggested by the referee, we extend out analysis to the Y(R') values provided by Gratton et al.\,(2010).

\section{Relationship between HB-morphology and globular cluster parameters.}
\label{results}
In this Section we investigate the correlations among $L1$, $L2$,
 and several physical and morphological parameters of their host GCs. Specifically, relations with metallicity, absolute  magnitude, and age are discussed in Sect.~\ref{sub:FeH},~\ref{sub:Mv}, and \ref{sub:Age} respectively.
 Section~\ref{sec:elio} investigates  the correlations with the internal variations of  the light elements and helium, while relations between $L1$ and $L2$ and the other parameters introduced in Sect.~\ref{GCpar} are analysed in Sect.~\ref{sub:altri}.

When we compare two variables, as we do in the next section for $L1$, $L2$, and [Fe/H], we use the Spearman's rank correlation coefficient $r$, to estimate the statistical dependence between the two. 
Uncertainties  in $r$ are estimated by means of bootstrapping statistics. We generated 1,000 resamples of the observed dataset, of equal size, and for each resample (i), (which is generated by random sampling with replacement from the original dataset) we estimated $r_{\rm i}$.
 We considered the dispersion of the $r_{\rm i}$ measurements ($\sigma_{\rm r}$) as indicative of robustness of $r$ and provide the number of  included GCs ($N$). 
\subsection{Metallicity}
\label{sub:FeH}
   \begin{figure*}[ht!]
   \centering
   \epsscale{.75}
      \plotone{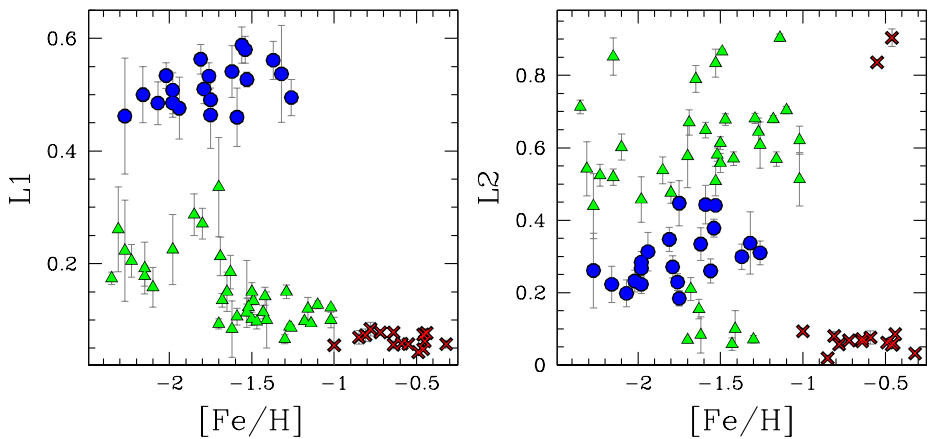}
      \caption{\textit{Left panel:} $L1$ versus GC metallicity for the 72 GCs studied in this work.  The G1, G2, and G3 GCs are colored red, green, and blue, respectively. \textit{Right panel:} $L2$ against metallicity.}
         \label{L12vsFe}
   \end{figure*}
The left panel of Fig.~\ref{L12vsFe} shows $L1$ against the GC metallicity.
An inspection of this plot reveals that all the metal-rich GCs have small $L1$ values and, hence, red-HBs. 
 At lower metallicities, there are GCs with almost the same iron abundance and yet different $L1$ values\footnote{The fact that $L1$ possesses a dependence on metallicity is expected from theory. Similarly, it is well known from theory that several other parameters may determine HB morphology and hence can affect the values of $L1$ and $L2$. This paper is an attempt to investigate those parameters that are actually at work by following a fully-empirical approach.}. This reflects the basic 2ndP phenomenon.
 Indeed, if the all of the GCs followed the same relation in the $L1$ versus [Fe/H] plane, we would assume that metallicity alone is sufficient to determine $L1$. The fact that we observe clusters with the same [Fe/H] but different $L1$ values, indicates that, apart from metallicity, at least one more parameter is at work.

 Our finding that the analysed GCs populate distinct regions  in the $L1$ versus [Fe/H] plane, and that the 2ndP phenomenon is absent among the majority of metal-rich GCs motivated us to define three groups of GCs as follows:
\begin{enumerate}
\item The first group, G1, includes all the metal-rich GCs ([Fe/H]$\ge-$1.0); 
\item the second, G2, is made of GCs with [Fe/H]$<-$1.0 and $L1\le$0.4; and
\item the remaining GCs with $L1>$0.4 belong to G3. 
\end{enumerate}
 Since the 2ndP phenomenon is absent among the majority of G1 GCs,
we will also consider a group that includes all the GCs in G2 and G3, hereafter G2+G3. 
The statistical analysis presented in the following will be provided for all the GCs together, as well as for the different groups separately.

There is a significant anticorrelation between $L1$ and [Fe/H] among G2 GCs, with a Spearman's rank correlation coefficient $r_{\rm G2}=-0.70$ ($\sigma_{\rm r, G2}=0.08$, $N_{\rm G2}=38$) that drops to $-0.88$ if we consider G1 and G2 GCs together, an even stronger correlation.
The Spearman's correlation coefficients and the corresponding $\sigma_{\rm r}$ values are listed in Tab.~3 for the 
groups of GCs defined above. 
There is no correlation between $L2$ and [Fe/H], either for G1 or G3 GCs. 

 Apart from NGC\,6388 and NGC\,6441, all G1 GCs host a purely red HB and have $L2$ values smaller than the majority of the other GCs.  
In G2 and G3 GCs metallicity is not responsible for the extension of the HB ($L2$)  as shown in the right panel of Fig.~\ref{L12vsFe}.

\subsection{Absolute Magnitude}
\label{sub:Mv}
   \begin{figure*}[ht!]
   \centering
   \epsscale{.75}
      \plotone{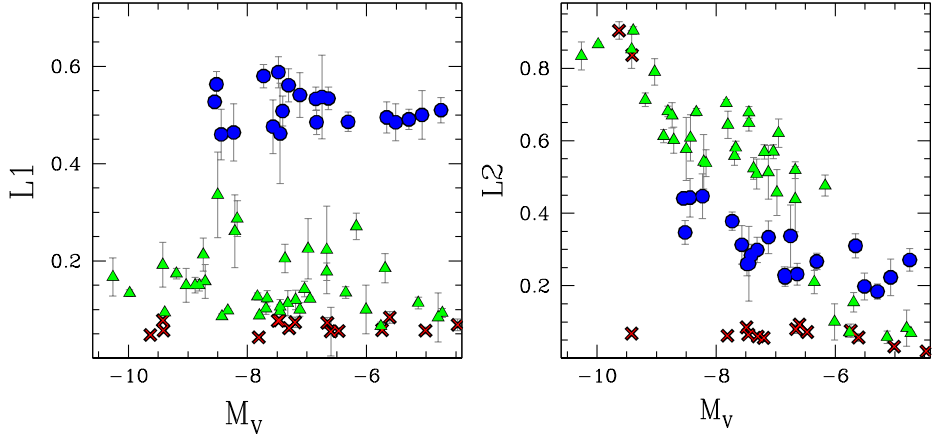}
      \caption{ $L1$ (left panel) and $L2$ (right panel) as a function of GC absolute  magnitude.}
         \label{L12vsMv}
   \end{figure*}
The left panel of Fig.~\ref{L12vsMv} shows that there is no significant correlation between  $L1$ and the GC absolute luminosity for any of the  
groups of GCs defined above.
In contrast, both G2 and G3 GCs exhibit significant anti-correlations between $L2$ and the absolute GC magnitude, which relates to the GC mass assuming all GCs have roughly the same mass-to-light ratio. This is shown in the right panel of Fig.~\ref{L12vsMv}, where we plot $L2$ as a function of $M_{\rm V}$. The Spearman's rank correlation coefficient is $r_{\rm G2}$=$-$0.89 ($\sigma_{\rm r, G2}=0.05$, $N=38$), and  $r_{\rm G3}$=$-$0.71 ($\sigma_{\rm r, G3}=0.13$, $N_{\rm G3}=21$) for the G2 and G3 samples, respectively, $r_{\rm G2+G3}$=$-0.80$ ($\sigma_{\rm r, G2+G3}=0.06$, $N_{\rm G2+G3}=59$) for G2+G3 GCs.

\subsection{Age}
\label{sub:Age}
\begin{figure*}[htp!]
   \epsscale{.45}
   \plotone{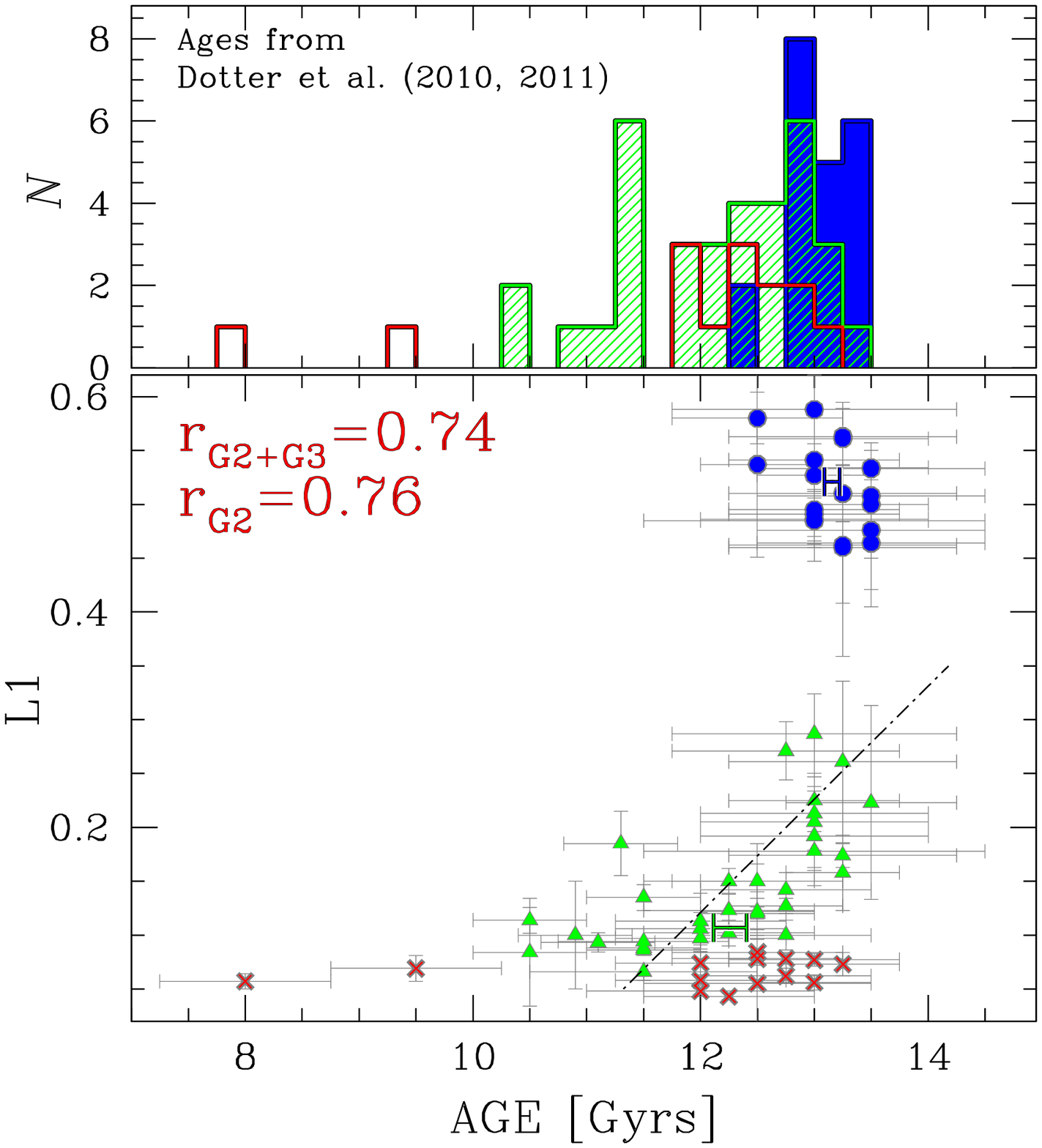}
   \plotone{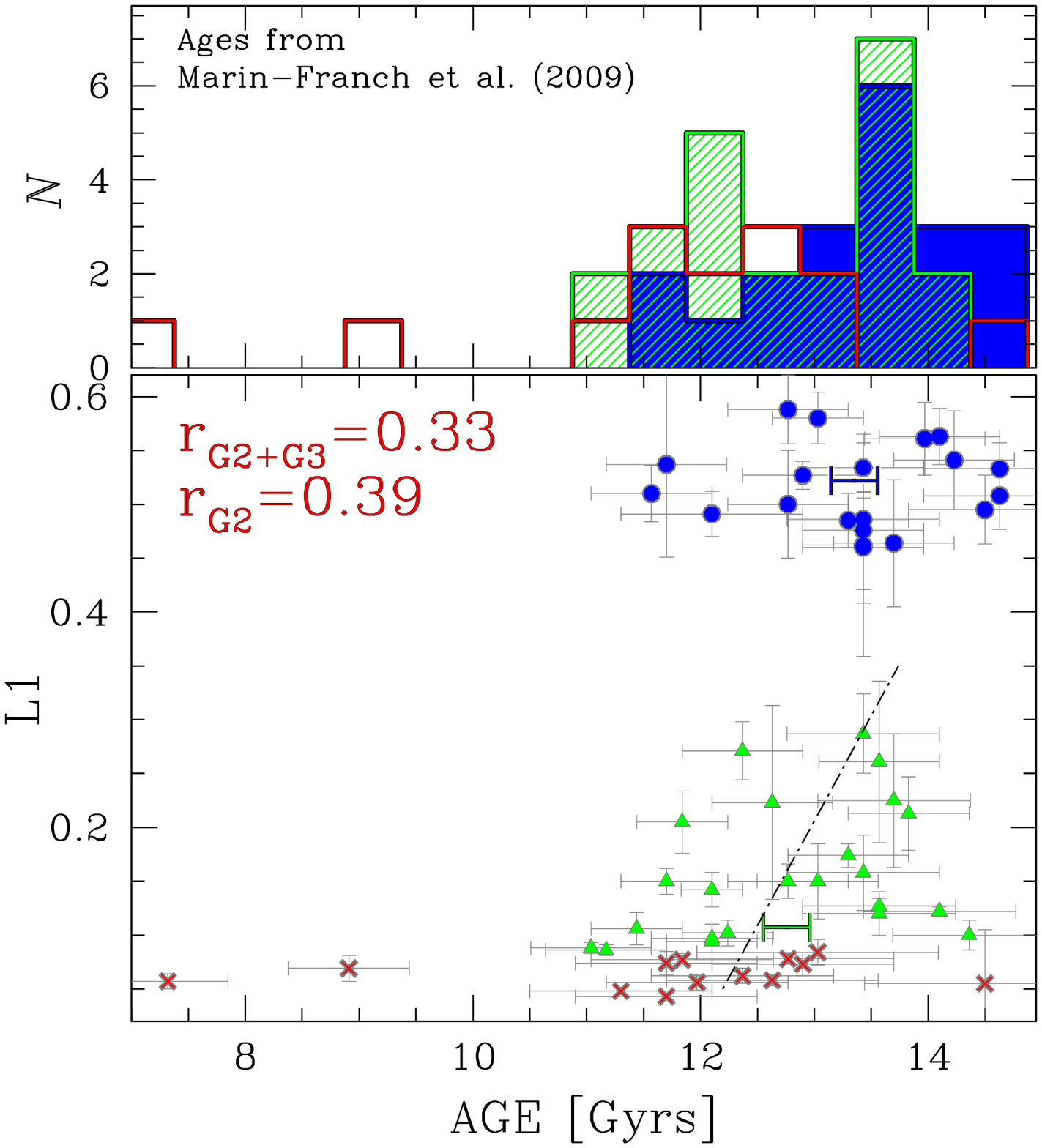}
   \plotone{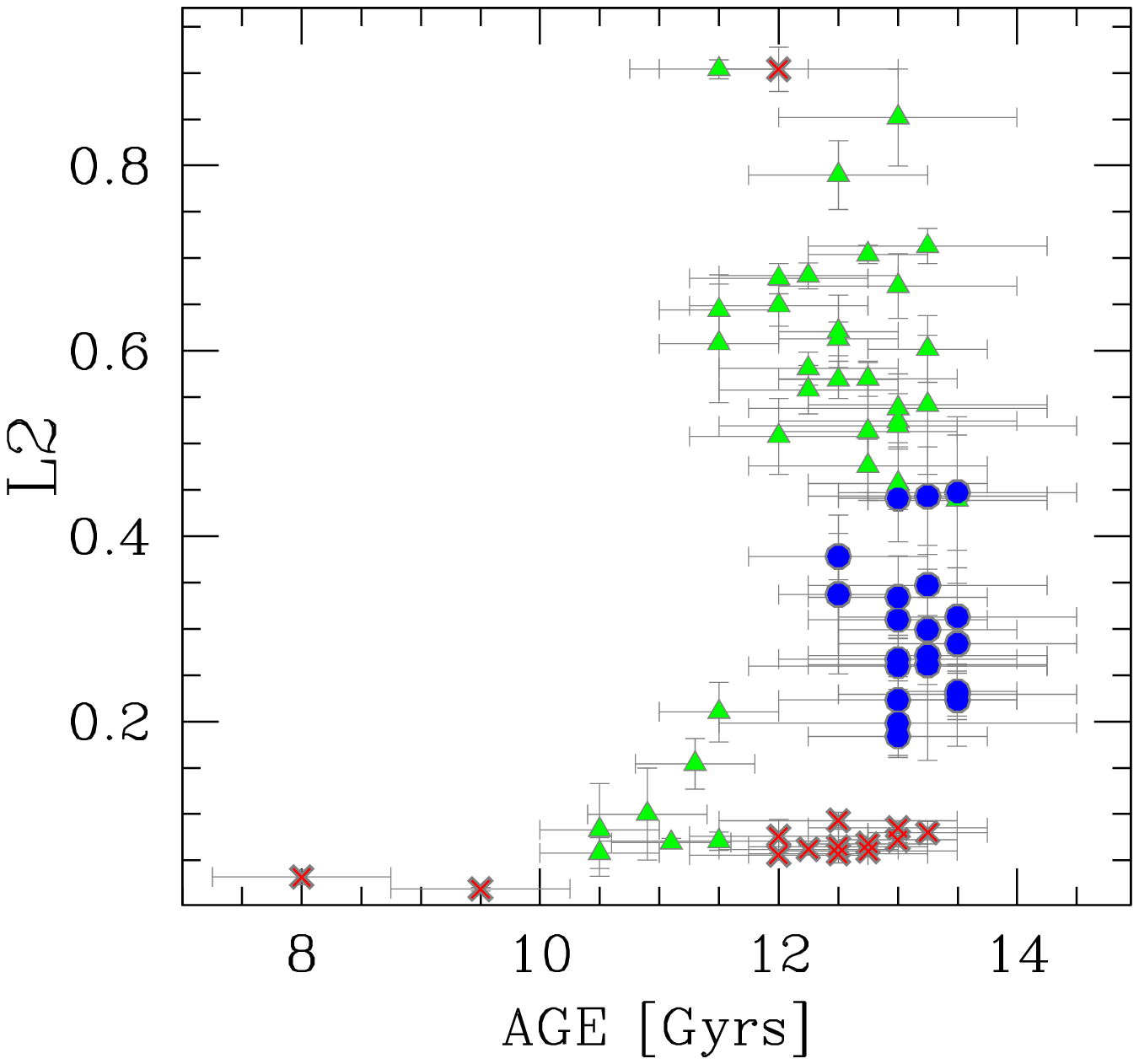}
   \plotone{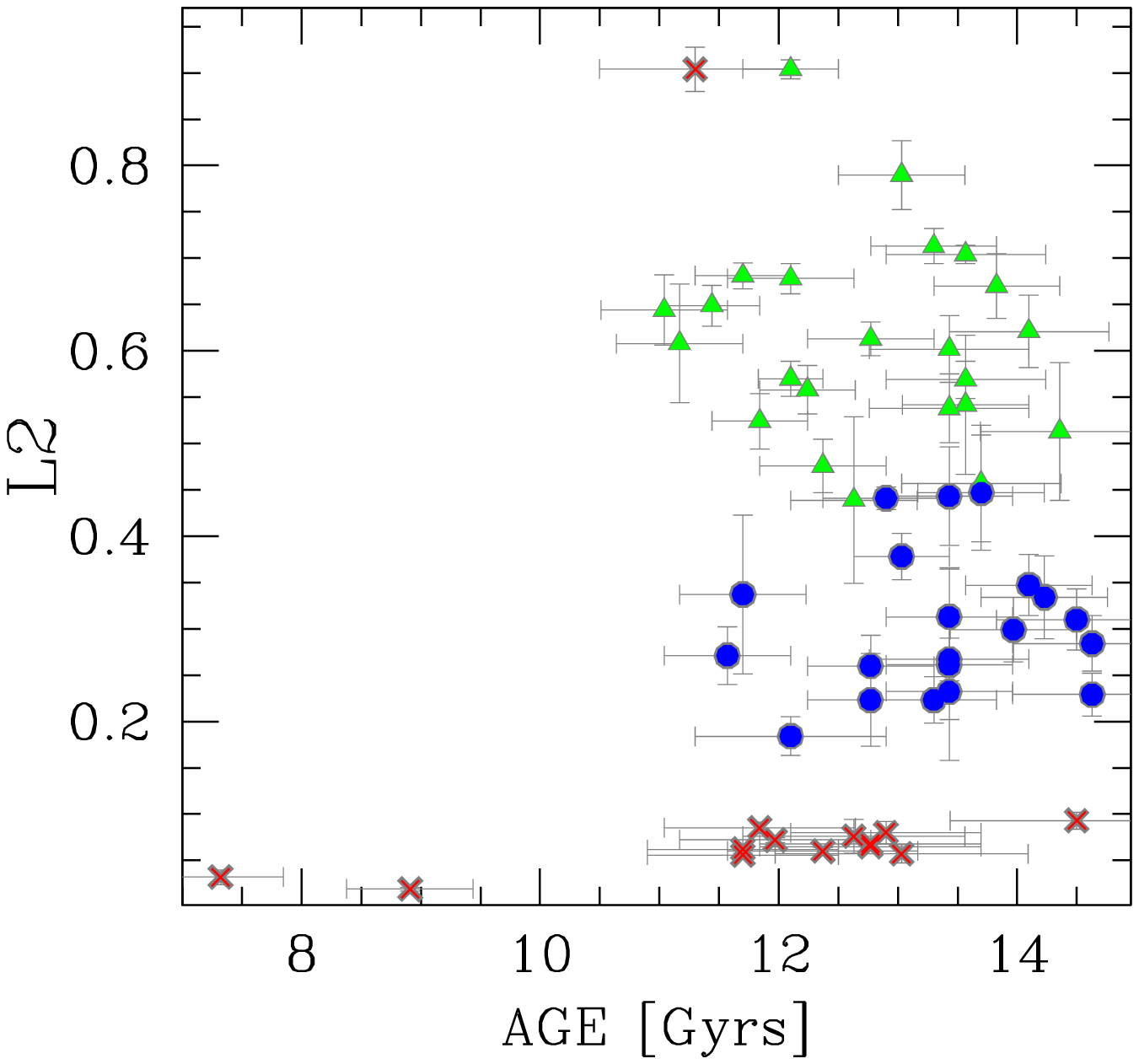}
\caption{
 $L2$ (lower panels) and $L1$  (middle panels) against age. The histograms of age distribution for the G1 (red), G2 (green), and G3 (blue) GCs are plotted in the upper panel.
 In the left and right panels we used age measurements from  Dotter et al.\,(2010, 2011) and Mar{\'{\i}}n-Franch et al.\,(2009), respectively. Black dashed-dotted  lines  in the middle panel are the best-fitting straight lines for the G2 sample. The Spearman's coefficients $r_{\rm G2}$ and $r_{\rm G2+G3}$ are also indicated.}
\label{LvsAge1}
\end{figure*}

\begin{figure*}[htp!]
   \epsscale{.45}
   \plotone{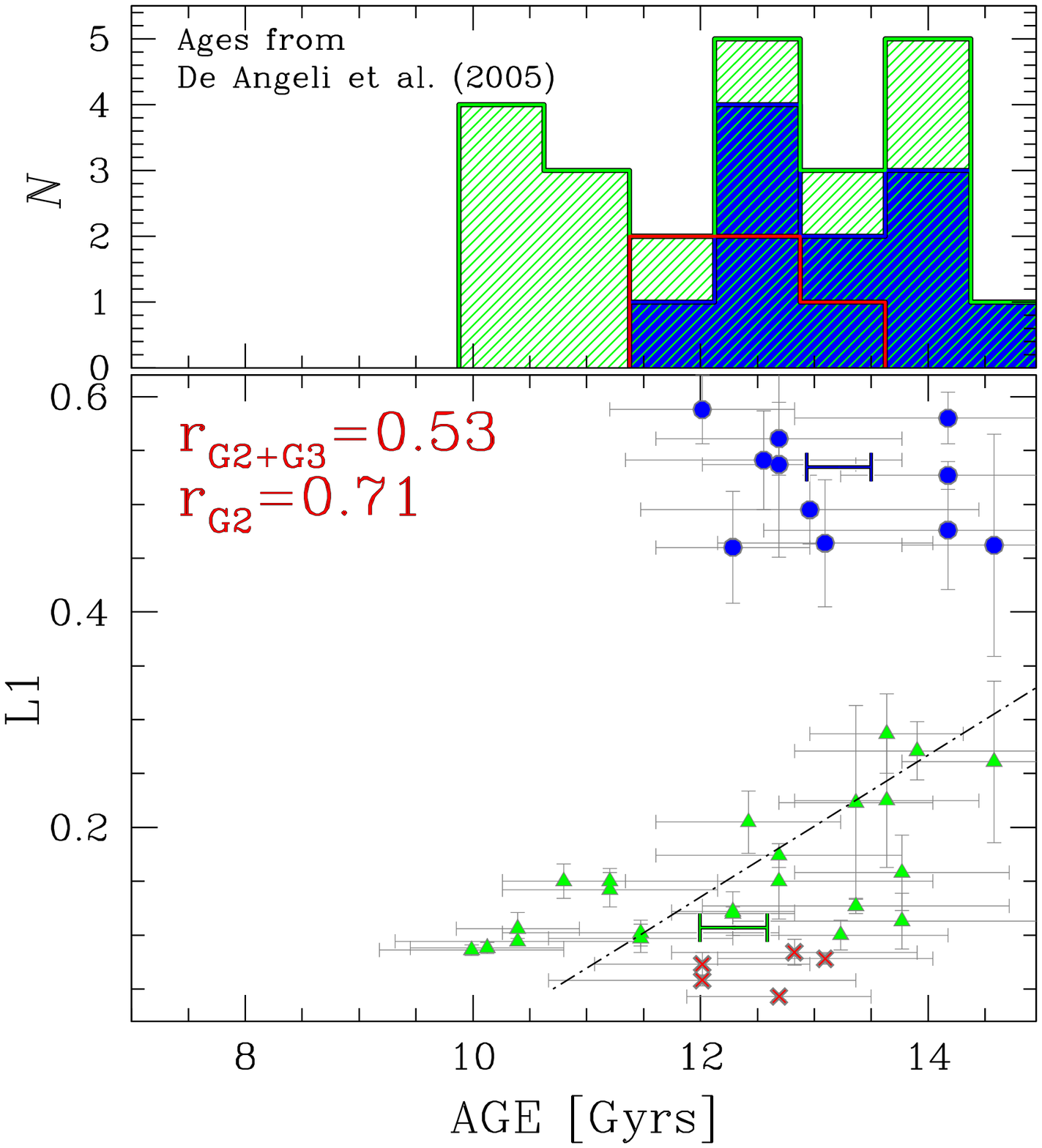}
   \plotone{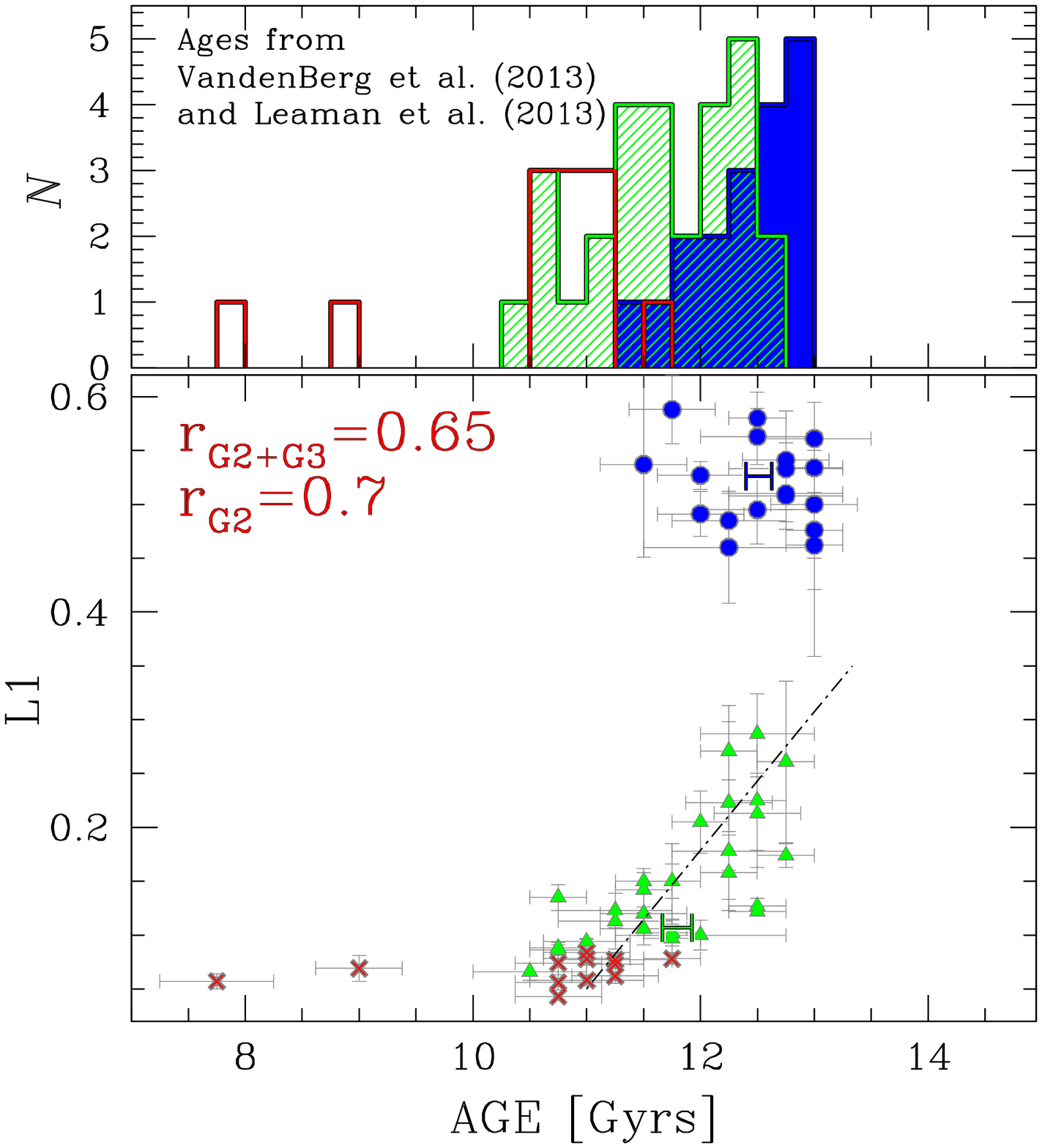}
   \plotone{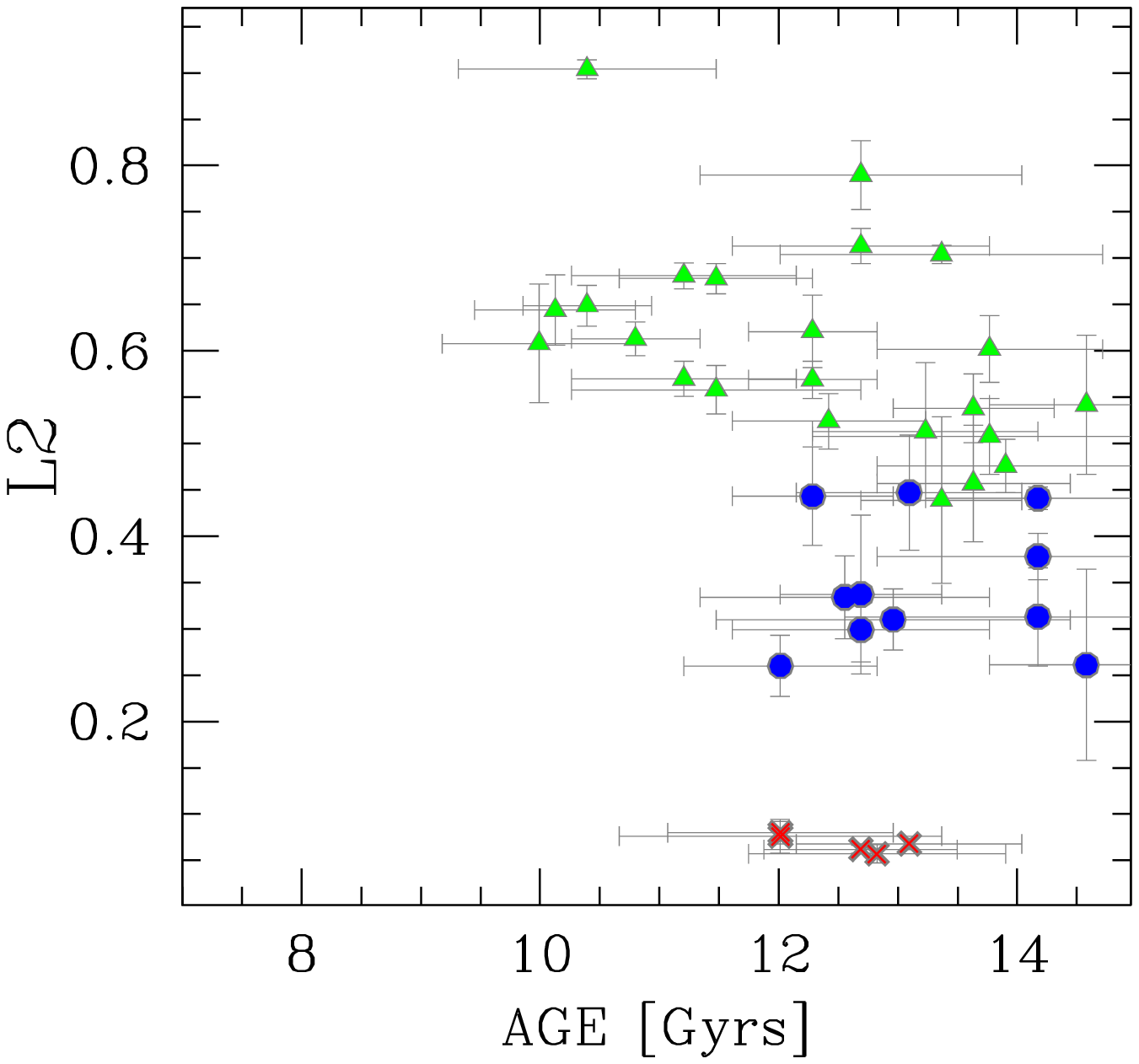}
   \plotone{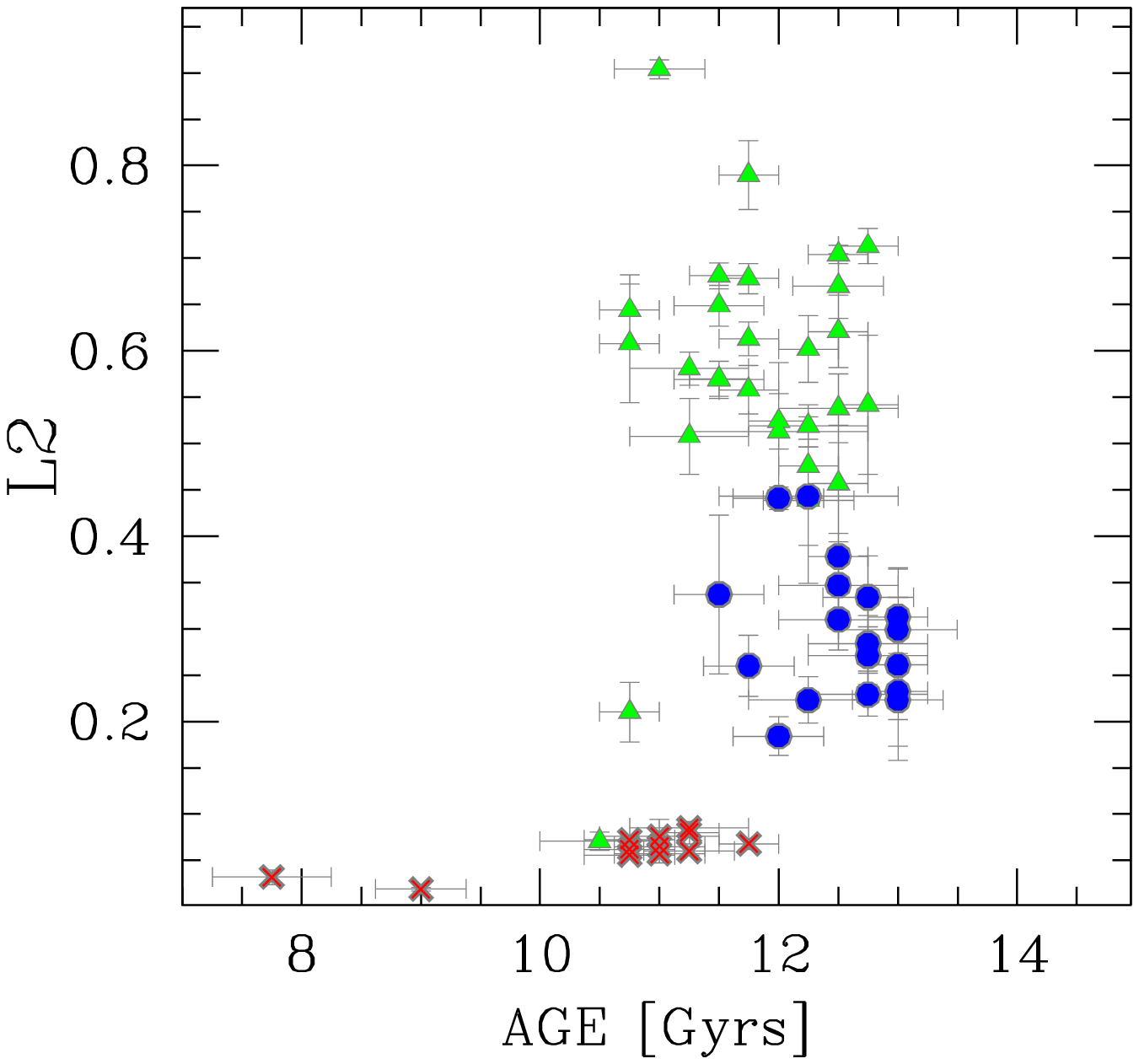}
\caption{As in Fig.~\ref{LvsAge1}, but for age measurements from De Angeli et al.\,(2005, left panels) and for ages derived by VandenBerg et al.\,(2013) and Leaman et al.\,(2013, right panels).}
\label{LvsAge2}
\end{figure*}

Histograms of the age distributions for G1, G2, and G3 GCs are shown in the upper panels of Fig.~\ref{LvsAge1} for the age measurements from Dotter et al.\,(2010, 2011, left panel)
and Mar{\'{\i}}n-Franch et al.\,(2009, right panel), and in Fig.~\ref{LvsAge2} for the age measurements of De Angeli et al.\,(2005, left panel)
and VandenBerg et al.\,(2013) and Leaman et al.\,(2013, right panel). On average, G3 GCs are systematically older than G2 GCs, and this result is independent of the adopted age scale.
Specifically, independently from the four adopted age scales,
on average, G2 GCs are younger than G3 ones by $\sim$1 Gyr with G3 GCs clustered around the value of $\sim$13 Gyrs, and G2 GCs spanning a wider age interval. The mean ages of G2 and G3 GCs are listed in Tab.~4.

$L1$ is plotted as a function of GC age in the middle panels of Figs.~\ref{LvsAge1} and~\ref{LvsAge2}.  
There is a  positive correlation between age and $L1$ for G2 GCs, with older G2 GCs having, on average, greater $L1$ values. The Spearman coefficient is high $r_{\rm G2}\geq0.70$, except when we adopt ages from  Mar{\'{\i}}n-Franch et al.\,(2009) indicating that in the latter case the significance level is low\footnote{As pointed out by the referee, there are three G3 GCs which, according to  Mar{\'{\i}}n-Franch et al.\,(2009) are younger than $\sim$12.25 Gyrs. Their ages are in agreement with the average age of G3 GCs only to within 1.5-2.5 $\sigma$.  We are not able to say if this difference is due to measurement errors or is intrinsic.}. 

As a further check we have divided GCs into three sub-groups with almost the same metallicity. We have defined a metal poor ([Fe/H]$<$$-$1.7), a metal intermediate ($-$1.7$<$[Fe/H]$<$-1.4), and a metal rich ([Fe/H]$>$$-$1.4) group, and investigate age-$L1$ relation for GCs in each of them. Results are listed in Tab.~4. In all the cases, G3 GCs are systematically older even if, especially for metal-poor GCs, the statistical significance of the measured age difference is marginal, but we are limited by small number statistics. 
The fact that G3 GCs are systematically older than G2 GCs and the presence of significant correlation between age and $L1$ for G2 GCs indicate that GC age is partly responsible for the color distance between the RGB and the reddest part of the HB, being metallicity the other parameter for L1 extension\footnote{ Gratton et al.\,(2010) also derived ages for a subsample of the clusters studied in the present paper. To determine these ages, they used stellar masses derived from isochrones that include ages from  Mar{\'{\i}}n-Franch et al.\,(2009) and De Angeli et al.\,(2005) already analysed in this paper. In addition, the way these ages are derived is closely related to the HB morphology. For these reasons we prefer to avoid to use them in the context of this paper. Nevertheless, we verified that our conclusions remain unchanged when ages from Gratton et al.\,(2010) are used.}.
 
There is no significant correlation between $L2$ and age as shown in the lower panels of Fig.~\ref{LvsAge1} and~\ref{LvsAge2}.
GCs with a double or multimodal SGB, namely NGC\,1851, NGC\,6388, NGC\,6656, and NGC\,6715  have been excluded from the statistical analysis above. 
As discussed in Sect.~\ref{age}, the large fraction of faint-SGB stars observed in these GCs can affect the age measurements.
 For completeness we provide in Tab.~3 the values of the Spearman's rank correlation coefficients to estimate the statistical significance of $L1$, $L2$ and age  correlations, together with results for the whole sample of GCs. The main results of this section remain unchanged when the GCs above are included in the analysis.   

\subsection{Helium and light elements}
\label{sec:elio}
\begin{figure*}[ht!]
   \centering
   \epsscale{.75}
      \plotone{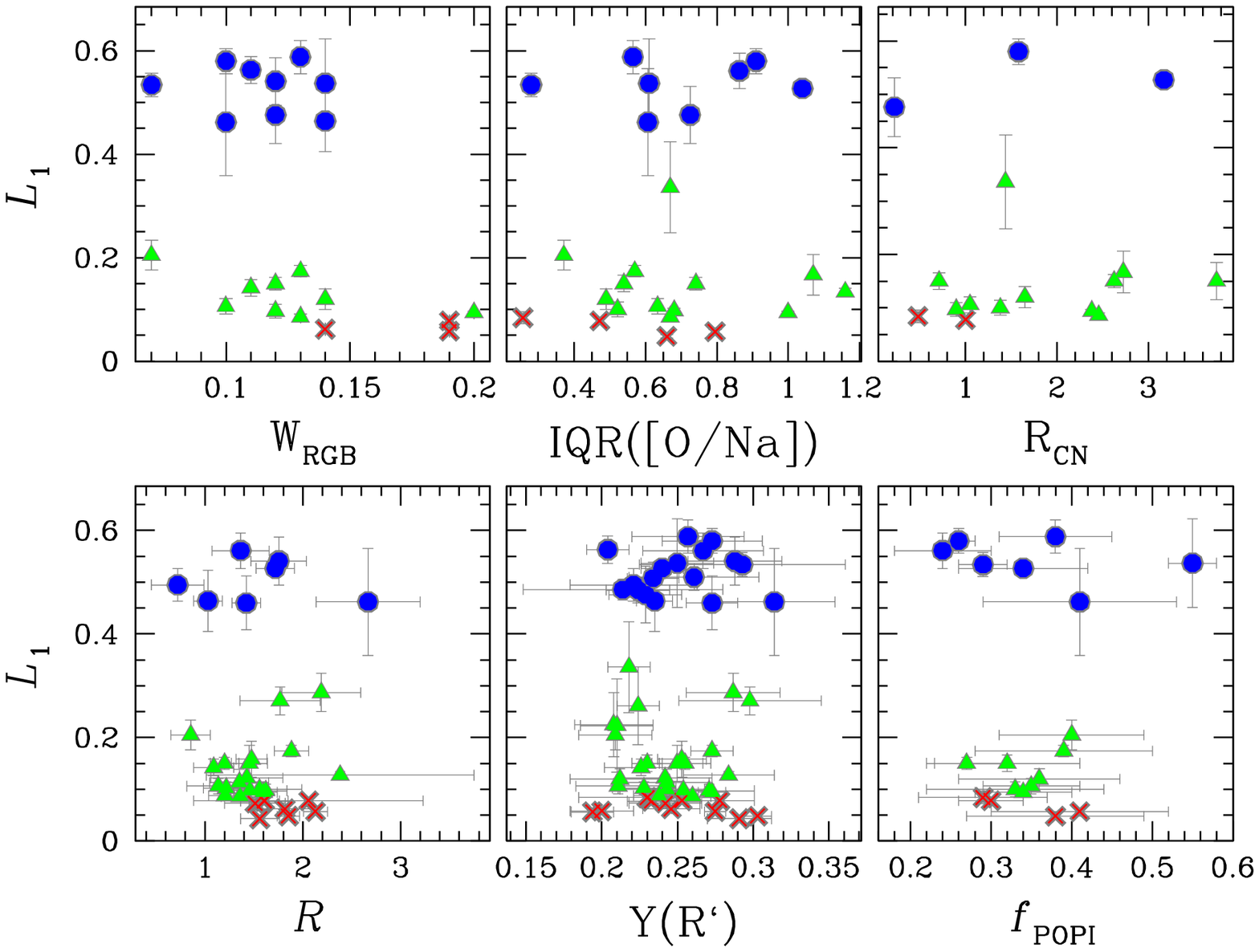}
      \caption{\textit{Upper panels}: $L1$ as a function of the width of the RGB in the $c_{\rm UBI}$ index (left), the interquartile range of [O/Na] ratio (middle), and  the ratio between CN-strong and CN-weak stars (right). 
\textit{Lower panels}: $L1$ as a function of the R-parameter, that is the number ratio of HB to RGB stars brighter than the HB level (left),  the helium abundance inferred by Gratton et al.\,(2010) from the Iben\,(1968)  method based on the R-parameter (middle), and the fraction of first-population stars (right).}
         \label{L1vsChem}
   \end{figure*}

\begin{figure*}[ht!]
   \centering
   \epsscale{.75}
      \plotone{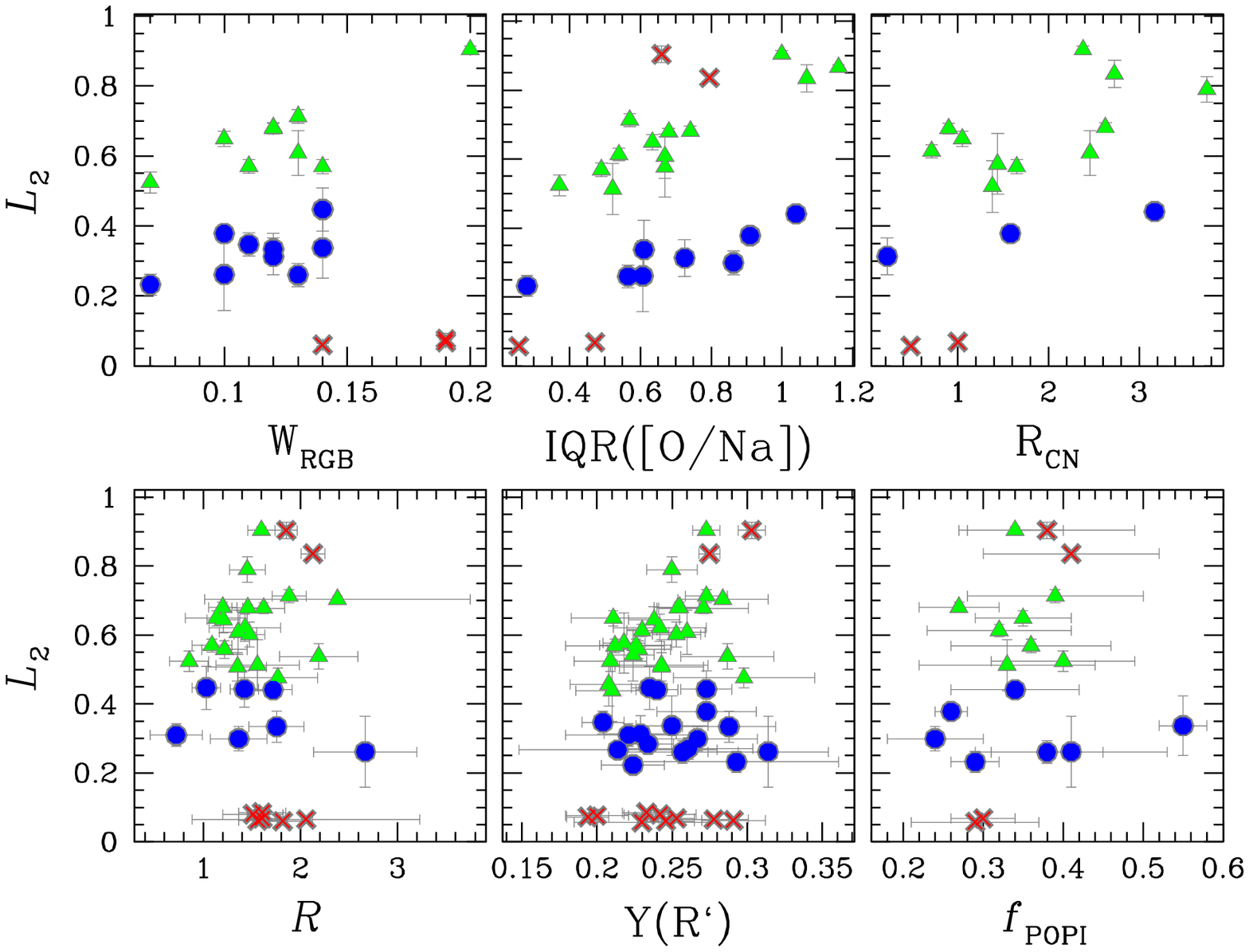}
      \caption{ As in Fig.~\ref{L1vsChem}, but for $L2$. }
         \label{L2vsChem}
   \end{figure*}

\begin{figure*}[ht!]
   \centering
   \epsscale{.75}
      \plotone{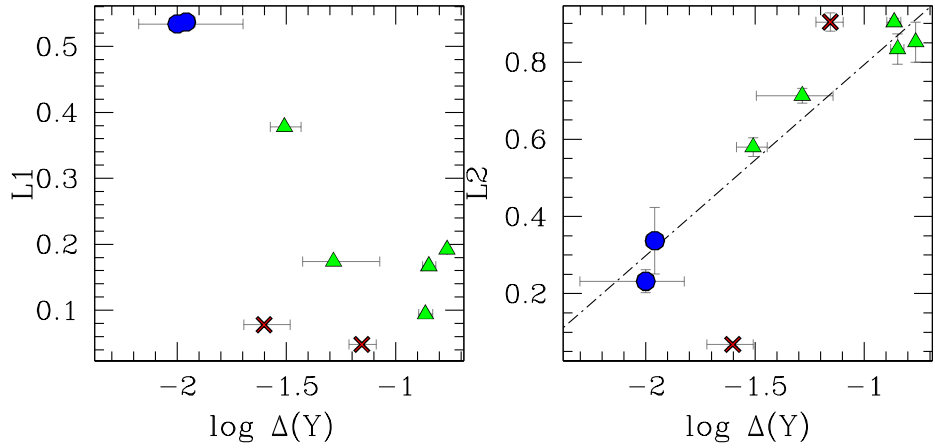}
      \caption{$L1$ (left panel) and $L2$ (right panel) as a function of the logarithm of the maximum helium difference among stellar populations in GCs. The black line is the best-fitting straight line for G2+G3 GCs.}
         \label{L2vsHe}
   \end{figure*}

As mentioned in Sect.~\ref{introduction}, the recent findings that in some GCs groups of stars with different light element abundances populate different HB segments strongly suggest that certain aspects of HB morphology may be strictly connected with multiple populations.
 To further investigate this scenario, in Figs.~\ref{L1vsChem},~\ref{L2vsChem},  and~\ref{L2vsHe} we show the relations between $L1$ and $L2$ and those quantities indicating  intracluster chemical variations, which we introduced in Sect.~\ref{elio}. 
As shown in Figs.~\ref{L1vsChem},~\ref{L2vsChem}, there is no significant correlation between $L1$ and $W_{\rm RGB}$, IQR([O/Na]), or $R_{\rm CN}$, $f_{\rm POPI}$, nor between $L2$ and $W_{\rm RGB}$, and $f_{\rm POPI}$.
 No significant correlations are observed between $L1$, $L2$ and $R$ or Y(R'). A mild correlation between $L2$ and Y(R') can not be ruled out for G2 GCs ($r_{\rm G2}$=0.44, $\sigma_{r, G2}$=0.21).
  
 Figure~\ref{L2vsChem} also shows that on average GCs with extended HBs have more extended Na-O anticorrelations, as demonstrated by the significant correlation between $L2$ and IQR([O/Na]) obtained for GCs in both G2 and G3.
 This result confirms the findings by Carretta et al.\,(2007) and Gratton et al.\,(2010).  
Among G2 GCs, those with large CN-strong and CN-weak populations ($R_{\rm CN}>2$) have, on average a more extended HB. 
The small number of G2 and G3 GCs where $R_{\rm CN}$ measurements are available prevents us from making any strong conclusion on the significance of the correlation with $L1$ and $L2$.

Theoretical models predict that star-to-star light-element variations observed in GC stars are associated with helium differences  that lead to HB stars with different masses because of the well-known inverse relationship between helium abundance and stellar mass for fixed metallicity and age (e.g.\,Ventura \& D'Antona\,2005 and references therein).\footnote{The helium content of a star affects its location along the HB as follows. When compared with helium-normal stars (Y$\sim0.25$) of the same age and metallicity, He-enhanced stars have shorter main sequence lifetimes, and hence smaller masses at the main sequence turn-off for a fixed age. Assuming the He-normal and He-enhanced stars lose a similar amount of mass on the RGB, the He-enriched stars will have a smaller H-rich envelope on the HB and, thus, populate a hotter, bluer portion of the HB than the He-normal stars (e.g.\,Cassisi et al.\,2012).} 
The relation between  $L1$, $L2$ and the maximum internal helium difference measured from MS studies is plotted in Fig.~\ref{L2vsHe}.
The tight correlation between $L2$ ($r_{\rm G2+G3}$=$0.89$, $\sigma_{\rm r, G2+G3}$=$0.17$, $N_{\rm G2+G3}$=7) and the small corresponding value of $\sigma_{\rm r}$ for G2+G3 clusters confirms theoretical indications that helium-enhanced stellar population are responsible of the HB extension. 

\subsection{Relationships with other parameters of the host globular clusters}
\label{sub:altri}
\begin{figure*}[ht!]
   \centering
   \epsscale{.9}
      \plotone{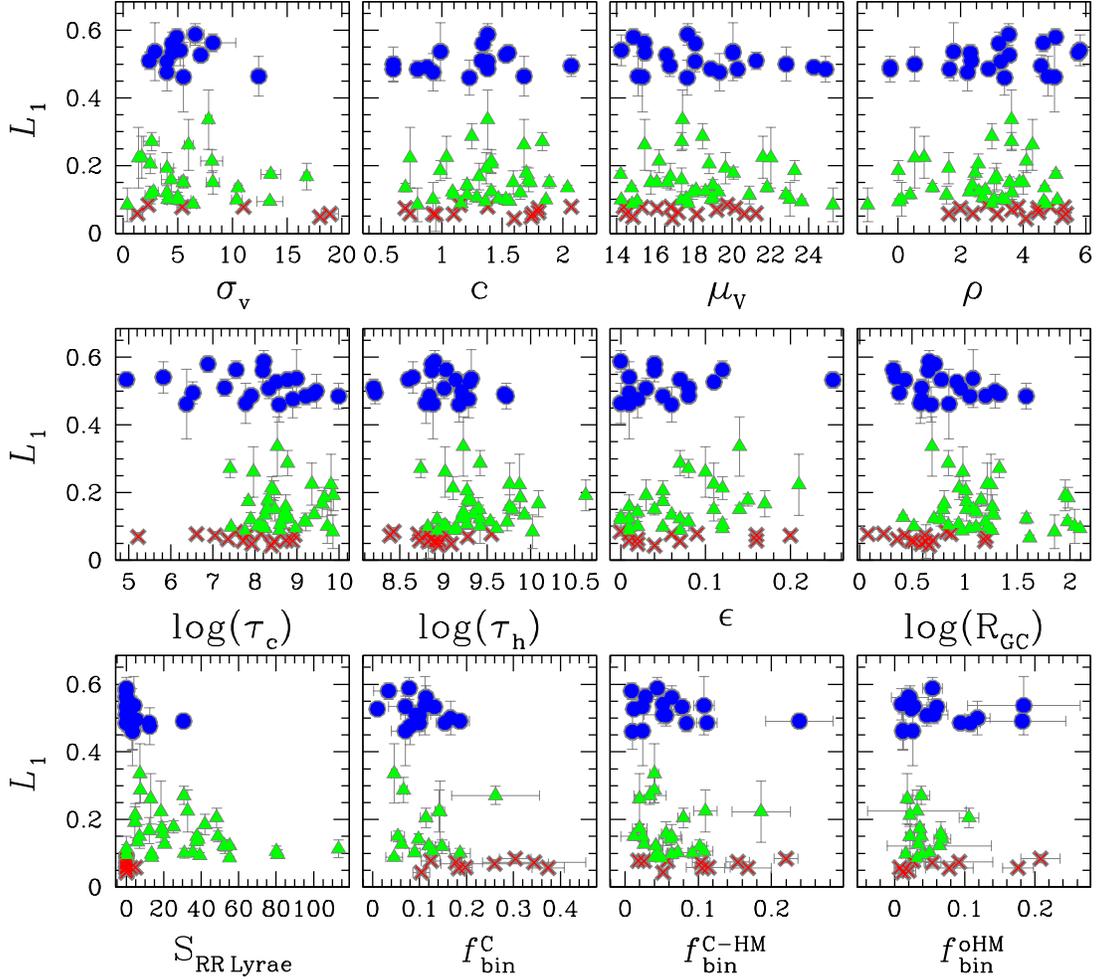}
      \caption{$L1$ as a function of some parameters of the host GCs.
     From the left to the right: central velocity dispersion, King\,(1962) model central concentration central luminosity brightness, and central luminosity (top) density, core and half mass relaxation time,  GC ellipticity, and logarithm of Galactocentric distance (middle), 
frequency of binaries, and fraction of binaries in the core, in the region between the core and the half-mass radius, and outside the half-mass radius (bottom).  $\sigma_{\rm v}$ is given in km~s$^{-1}$, $\mu_{\rm V}$ in $V$ magnitude per square arcsecond, $\tau_{\rm c}$ and $\tau_{\rm h}$ in years, and $R_{\rm GC}$ in kpc.
}
         \label{L1vsPar}
   \end{figure*}

Figure~\ref{L1vsPar} shows other monovariate relations  involving $L1$.
There is no significant correlation between $L1$ and
 central velocity dispersion, King\,(1962) model central concentration,
  central brightness, central density, core and half mass relaxation time, GC ellipticity, Galactocentric distance, and binary fraction.
This is confirmed by the values of the Spearman's correlation coefficient listed in Tab.~3. 

\begin{figure*}[ht!]
   \centering
   \epsscale{.9}
      \plotone{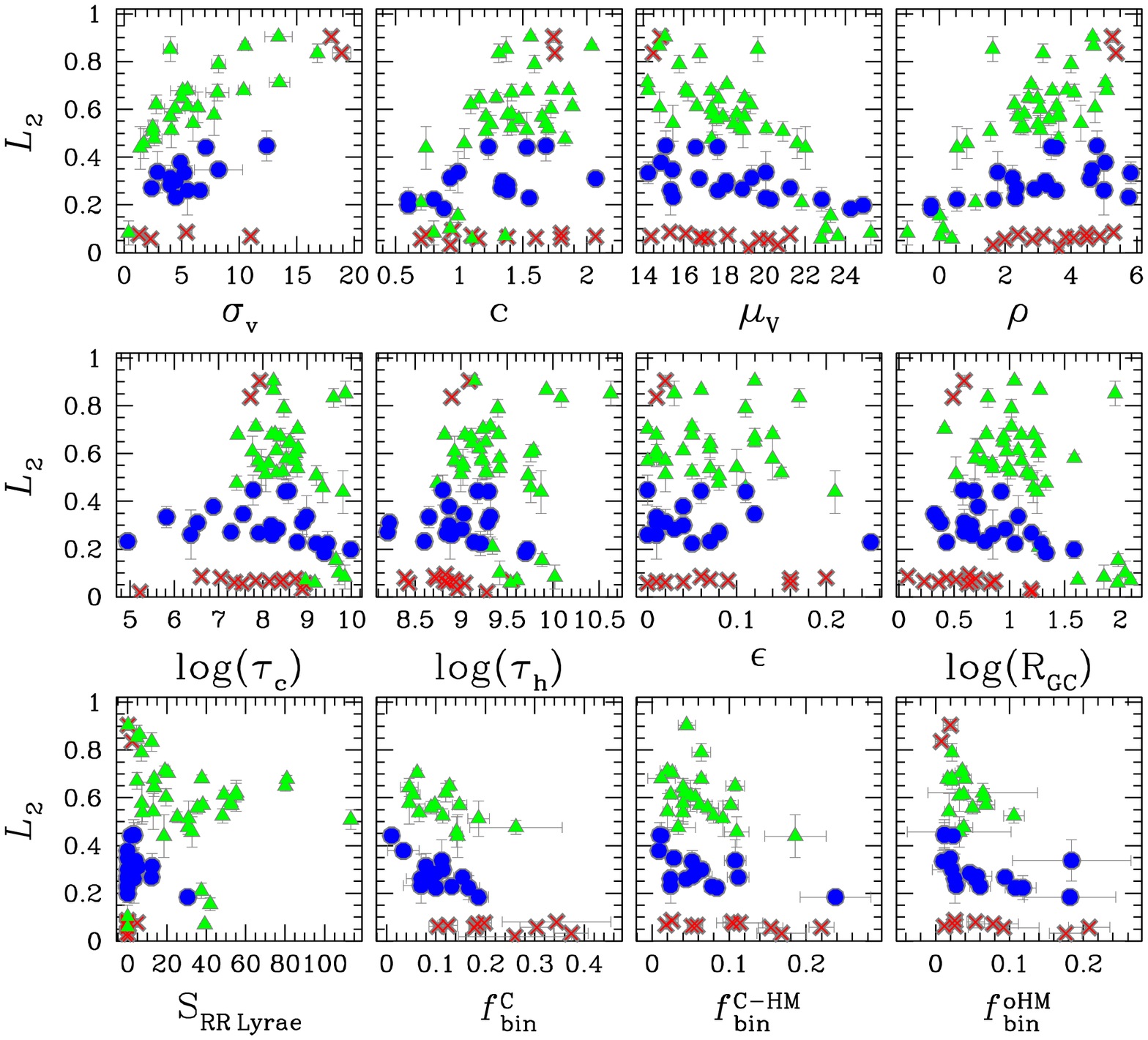}
      \caption{$L2$ as a function of the same parameters of the host GCs  as used in Fig.~\ref{L1vsPar}.}
         \label{L2vsPar}
   \end{figure*}

In Fig.~\ref{L2vsPar} we see that
$L2$ correlates with $\rho$ and $\sigma_{\rm V}$ for G2 and G3 GCs, and anticorrelates with $\mu_{\rm V}$ and $f_{\rm bin}^{\rm C, C-HM, oHM}$ for each group of GCs, even if the anticorrelation is less or not significant for the G2+G3 sample. These results are not unexpected as these quantities also correlate with GC mass 
 (Djorgovski \& Maylan\,1994).

\section{Comparison with the literature}
\label{literature}
 According to  Gratton et al.\,(2010), in the ADS data base there are more then two hundreds papers dedicated to the 2ndP phenomenon. 
Hence, any comparison here with the wide literature on the HB 2ndP can only be very far from complete.
In this section we  discuss some of the more relevant results. We refer the reader to review papers (e.g.\,Freeman \& Norris\,1981; Catelan\,2009) and references therein for a complete view on this topic.

\begin{figure*}[ht!]
   \centering
   \epsscale{.65}
      \plotone{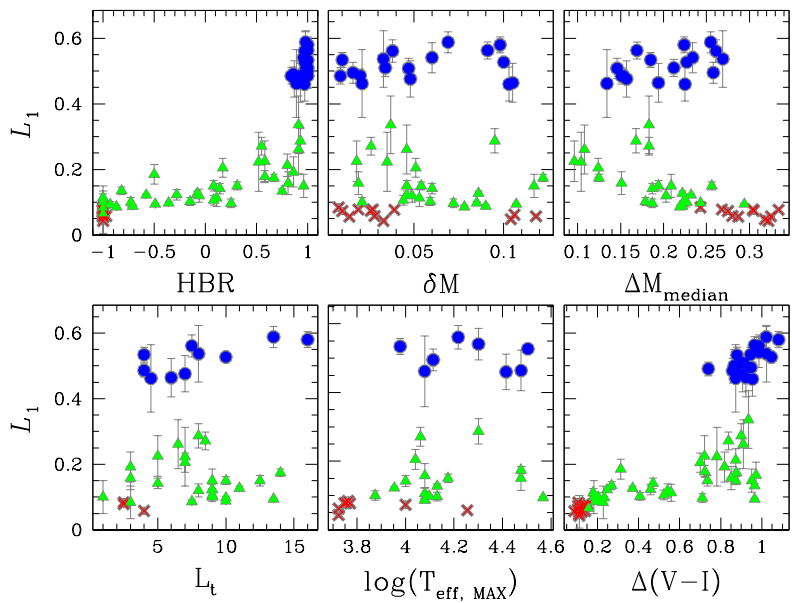}
      \caption{$L1$ as a function of several quantities used to parametrize HB morphology.  
Total length of the HB from Fusi-Pecci et al.\,(1993, lower-left panel) 
logarithm of the maximum temperature along the HB from Recio-Blanco et al.\,(2006, lower-middle panel),
 Median HB color ($\Delta$($V-I$)) from Dotter et al.\,(2010, lower-right panel), 
HBR ratio (upper-left panel),
maximum and median mass loss from 
 Gratton et al.\,(2010, upper-middle and upper-right panel).}
         \label{L1vslet}
   \end{figure*}

\begin{figure*}[ht!]
   \centering
   \epsscale{.65}
      \plotone{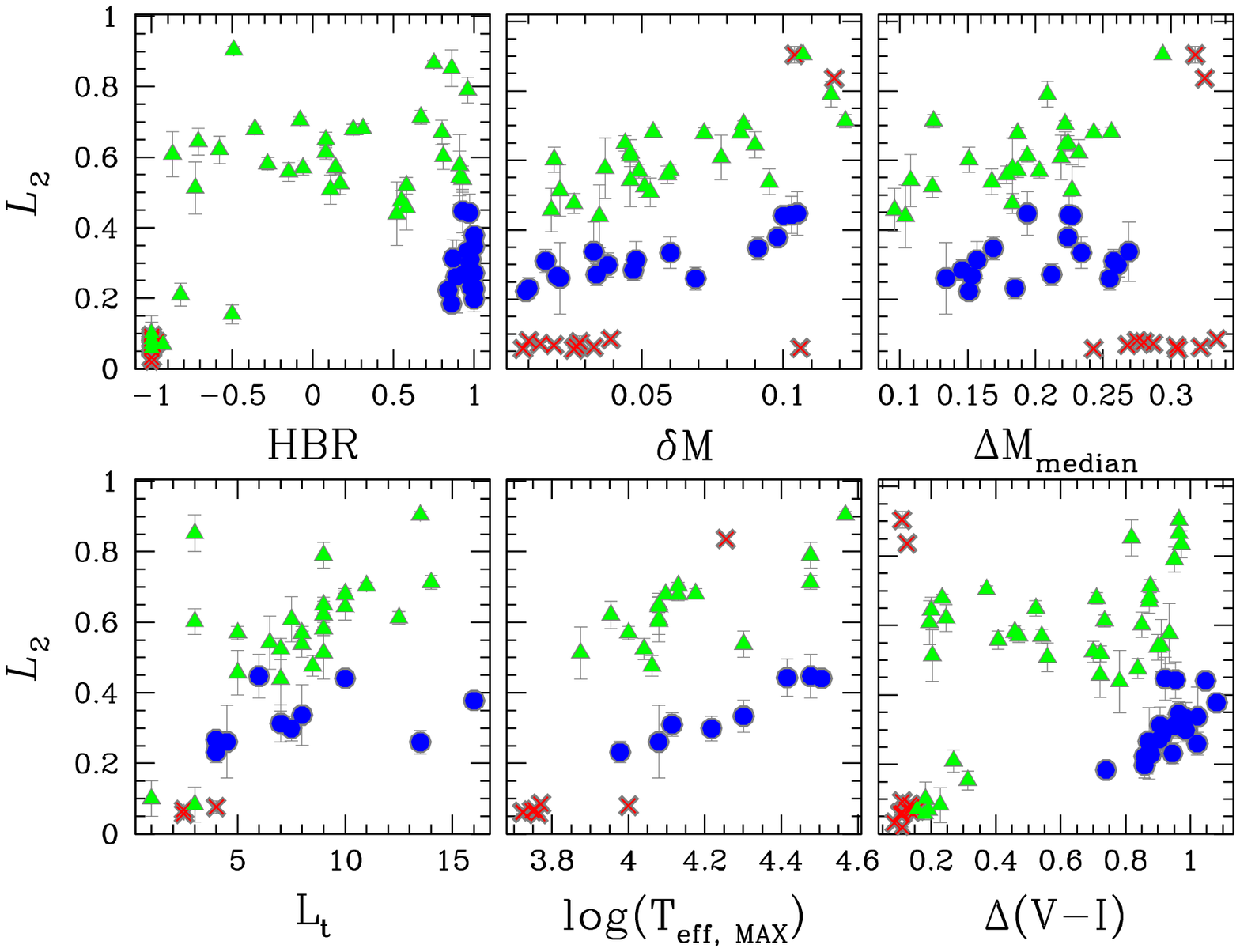}
      \caption{As in Fig.~\ref{L1vslet}, but for $L2$.}
         \label{L2vslet}
   \end{figure*}

As already mentioned in Sect.~\ref{introduction}, works on HB morphology in GCs  make use of different HB metrics.
In Fig.~\ref{L1vslet} and ~\ref{L2vslet}, we compare $L1$ and $L2$ with other quantities used to parametrize HB morphology.
The parameter to describe HB morphology that is mostly used in literature is the HB Type index or HBR (see Sect.~1).
Figures~\ref{L1vslet} and~\ref{L2vslet} compare $L1$ and $L2$ with HBR.  There is a linear correlation between $L1$ and HBR for G2 GCs, and then HBR saturates for G1 and for G3 GCs. $L2$ does not correlate with HBR.

 The wide literature on the 2ndP includes several works, similar to the present investigation, that are based on a fully-observational approach, together with others that also use a series of theoretical assumptions. A recent example of the latter is the paper by 
 Gratton et al.\,(2010), in which the authors
 used {\it HST}/WFPC2 and ground-based photometry of about one hundred GCs to derive median and extreme colours, and magnitudes of stars along the HB.  They used isochrones and horizontal branch evolutionary models to transform these colors into median and extreme masses of stars on the HB and adopted the median mass loss ($\Delta M_{\rm median}$=$M_{\rm RGB}-M_{\rm median}$) and the difference between the median and the minimum HB masses ($\delta$M=$M_{\rm median}-M_{\rm min}$) (where $M_{\rm RGB}$, $M_{\rm median}$, and $M_{\rm min}$ are RGB, median, and minimum HB masses, respectively) as parameters of the HB morphology. 
 To determine $\Delta M_{\rm median}$ and $\delta$M, Gratton et al.\,(2010) 
 assumed for each cluster the value of metallicity and age from Carretta et al.\,(2009). 
They find that the median mass loss correlates with metallicity and suggest  that, if the mass-loss law they used is universal, age is the 2ndP. They conclude that age  can explain the behaviour of the median HB when it is coupled with a given mass-loss law that is a linear function of [Fe/H]. 
 Gratton et al.\ also suggest that at least another parameter is needed to explain the HB morphology in GCs and argue that He abundance is the most likely candidate. 
 They show that star-to-star helium variations, when combined with a small random quantity, can reproduce the HB morphology, thus supporting the results of other authors 
 (e.g.\, D'Andona et al.\,2002; D'Antona \& Caloi\,2008; Dalessandro et al.\,2013).
They find that the HB extension correlates with the interquartile of the Na-O anticorrelation, as previously noticed by the same group of authors  (Carretta et al.\,2007).
Figures~\ref{L1vslet} and~\ref{L2vslet} show that $\delta$M correlates with $L2$ in G2 and G3 GCs, while there is no significant correlation between $L2$ and $\Delta M_{\rm median}$. The relation between $L1$ and $\Delta M_{\rm median}$ is similar to that of $L1$ and [Fe/H]. This reflects the tight correlation between  $\Delta M_{\rm median}$ and metallicity.

Fusi-Pecci et al.\,(1993) 
analysed 53 GCs and found that the net length ($L_{\rm t}$) of the HB and the presence and extent of blue tails are correlated with the GC density and concentrations, with more concentrated or denser GCs having bluer and longer HB morphologies.
 A correlation between HB morphology and absolute magnitude has been also detected by Recio-Blanco et al.\,(2006) 
who analysed the CMDs of 54 GCs obtained from homogeneous {\it HST} Wide Field Planetary Camera 2 data (Piotto et al.\,2002) 
 and concluded that the maximum effective temperature ($T_{\rm eff, MAX}$) encountered along the HB correlates with $M_{\rm V}$, with more-luminous GCs having also more-extended HBs.

\begin{figure*}[ht!]
   \centering
   \epsscale{.4}
      \plotone{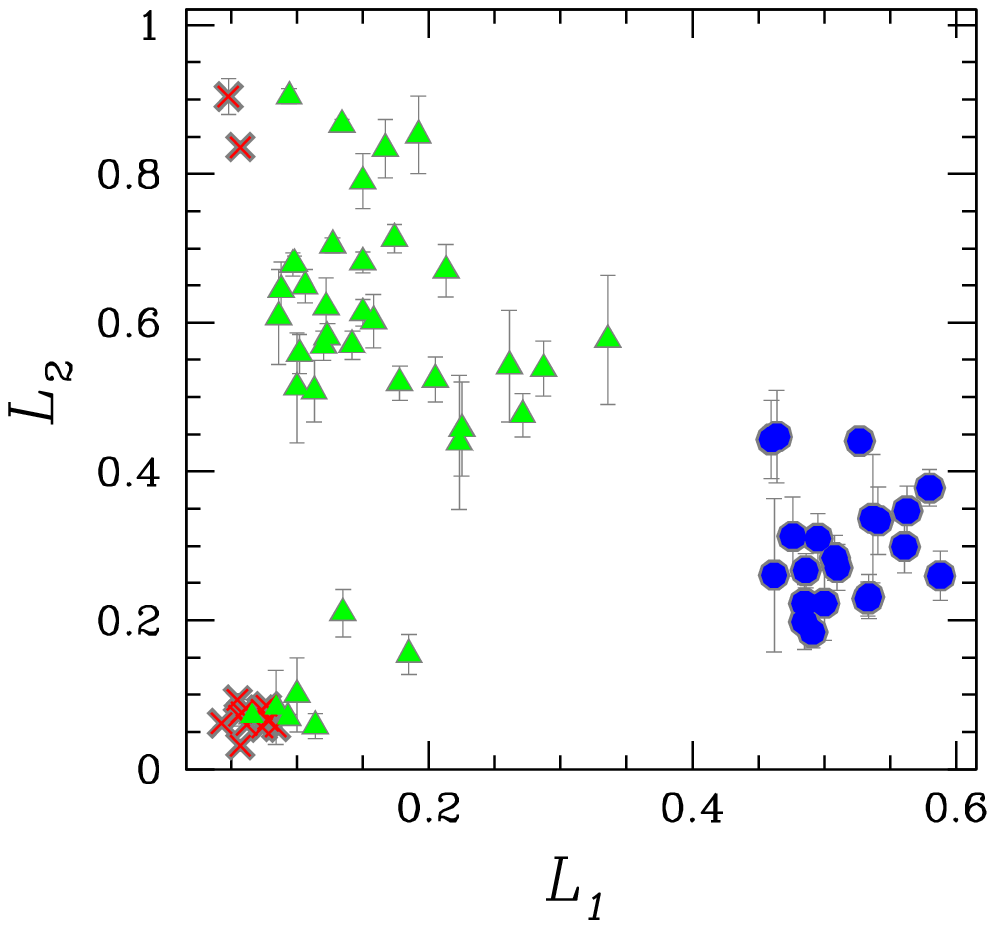}
      \caption{$L2$ as a function of $L1$ for the 74 GCs studied in this paper.}
         \label{L2vsL1}
   \end{figure*}

As discussed in Sect.~1, the way an HB morphology metric is defined influences the outcome. It explains why some studies conclude that mass and/or He content are the main driver of HB morphology, while others indicate age as the main 2ndP. 
The definition of the two parameters $L1$ and $L2$  and their comparison with other quantities commonly used to parametrize the HB morphology (like $L_{\rm t}$, $T_{\rm eff, MAX}$, and $\Delta$($V-I$)) may help to  shed some light on this controversy.
Figure~\ref{L1vslet} shows that  on the one hand there is no significant correlation between either $L_{\rm t}$ or $T_{\rm eff, MAX}$ and $L1$. On the other hand, among G2 GCs, $L1$ correlates with $\Delta$($V-I$) among G2 GCs and G3 GCs have, on average, larger $L1$ values than G2 GCs (indicative of older ages in G3 than G2). 
From Fig.~\ref{L2vslet} we note that $L2$ correlates with $T_{\rm eff, MAX}$ for both G2 and G3 GCs and a correlation between $L2$ and $L_{\rm t}$ is also observed for G2 GCs. Figure~\ref{L2vslet} reveals no significant correlation between $L2$ and $\Delta$(V$-$I). 
Finally, in Fig.~\ref{L2vsL1} we compare the two parameters introduced in this paper, and show that  $L2$ and $L1$ are not significantly correlated.

We conclude that both the metric defined by 
 Fusi-Pecci et al.\,(1993) and Recio-Blanco et al.\,(2006), 
as well as $L2$, are sensitive to some properties (possibly helium variations) of the HB morphology but lack sensitivity to others (such us age). In contrast $\Delta$($V-I$) and $L1$ are more sensitive to different properties of the HB (e.g.\, metallicity and age). The use of a pair of parameters, such as $L1$ and $L2$, can provide a more exhaustive description of the HB morphology than one alone.

\section{Summary and conclusions}
\label{summary}
In this paper we exploit both recent observational findings and ideas provided in the early 1980s to investigate the relation between HB morphology and various properties in GCs.
 These new findings come from studies on multiple stellar populations in GCs that show that the position of a star along the HB is connected to its chemical composition. First generation stars populate the cooler side of the HB and second generation (He-enriched) stars populate the hotter side. 

 Freeman \& Norris\,(1981) suggested that, apart from metallicity, at least two parameters are needed to explain the HB morphology. One of these should be a {\it global} parameter that varies from GC to GC, and the other a {\it non-global} parameter that varies within the GC.
Driven by this idea we defined two new parameters to describe the HB morphology: $L1$, which indicates the distance between the RGB and the coolest part or the HB, and $L2$, which measures the color extension of the HB.
Our analysis reveals that $L1$ depends on GC age and metallicity, while $L2$ correlates with the GC luminosity (hence, the mass) and the range of He content ($\Delta$Y).

These results suggest that, along the lines suggested by Freeman \& Norris, age and metallicity are the main {\it global} parameters of the HB morphology of GCs, while GC mass is related to the HB extension.
Works on multiple stellar populations in GCs show that more massive GCs exhibit, on average, larger internal helium variations, $\Delta$Y, than less massive GCs. $\Delta$Y is positively correlated with $L2$ and GC mass, though this analysis is limited to a small number of GCs at present. This makes it very tempting to suggest that internal star-to-star helium variation, associated with GC mass and the presence of multiple populations, is the main {\it non-global} parameter.\footnote{If this scenario is correct, in some metal-intermediate GCs with very small $L2$ values, like AM\,1, Eridanus, Pal\,3, Pal\,4, Pal\,14, Rup\,106, and Pyxis, any internal helium variation, if present, should be very small. These GCs are hence good candidates to host a simple stellar population.}
 The use of two quantities $L1$ and $L2$, that share a common definition (Sect.~3) but are sensitive to different phenomena allow us to discriminate the effects of global and non-global parameters on the HB morphology.\\

\begin{acknowledgements}
We thank F.\,D'Antona and J.\,Lattanzio for useful discussion and suggestion. 
We are gratefull to P.\,Stetson for providing unpublished light curves of RR\,Lyrae.  APM, JEN, and HJ acknowledge the financial support from the Australian Research 
Council through Discovery Project grant DP120100475.
 JEN is also supported by  the Australian Research Council through Discovery Project grant DP0984924.
AD acknowledges support from the Australian Research Council (grant FL110100012).
GP and SC acknowledge financial support from PRIN MIUR 2010-2011, project
 `The Chemical and Dynamical Evolution of the Milky Way and Local Group
 Galaxies', prot. 2010LY5N2T, and from PRIN-INAF 2011 `Multiple Populations in Globular Clusters: their role in the Galaxy assembly'.
Support for this work has been provided by the IAC (grant 310394), 
and the Education and Science Ministry of Spain (grants AYA2007-3E3506, and AYA2010-16717). 
\end{acknowledgements}

\appendix{{\bf \large A1. The impact of the RR\,Lyrae phase on $L1$ and $L2$ measurements.}\\

 To investigate the impact of excluding RR\,Lyrae that might be out of the instability strip in the determination of $L1$ and $L2$, we have simulated a number of CMDs for different choices of the fraction of RR\,Lyrae ($f_{\rm V}$), red-HB ($f_{\rm R}$), and blue-HB ($f_{\rm B}$) stars.
We assumed that RR\,Lyrae are distributed along the whole instability strip and  that all the RR\,Lyrae ab have the same light curve (this corresponds to the light-curve observed for V27 in M\,4, P$\sim$0.612d, and is one of the RR\,Lyrae ab in M\,4).  We assumed for all the RR\,Lyrae c the light curve of V40 in M\,4, P$\sim$0.299d. The reason why we have chosen these two RR\,Lyrae is that their amplitude and period are typical of RR\,Lyrae ab and c. 
The light curves have been kindly provided by Peter Stetson. They are based on more than 1,000 observations in {\it B}, {\it V}, and {\it R} bands and have been converted into F606W and F814W by using the color-temperature relations by 
 Dotter et al.\,(2008).
We choose V40, for which we have a light curve with the largest amplitude available to us, to maximize possible effects on the determination of $L1$ and $L2$.

 F606W and F814W magnitudes have been simulated at different phases to account for the fact that F606W and F814W images are taken at different times. We assumed $f_{\rm R}$=N$\times$($f_{\rm B}$+$f_{\rm R}$), (N=0.0, 0.04, 0.10, 0.50, 0.90, 0.96, 1.0); $f_{\rm V}$=M$\times$($f_{\rm B}$+$f_{\rm R}$+$f_{\rm V}$), (M=0,0.10,0.25,0.50) and simulated 1,000 CMDs for each combination of $f_{\rm B}$, $f_{\rm V}$, $f_{\rm R}$.
For each CMD we have calculated $L1_{\rm I}$, $L2_{\rm I}$ and $L1_{\rm II}$, $L2_{\rm II}$. These are the values of $L1$ and $L2$ obtained when all RR\,Lyrae lie within the instability strip, and when RR\,Lyrae are at random phase, respectively. 
The differences $\Delta L1=L1_{\rm I}-L1_{\rm II}$ and $\Delta L2=L2_{\rm I}-L2_{\rm II}$ are maximal in the case of a HB made of RR\,Lyrae only ($\Delta L1=$0.10, $\Delta L2=$0.22).  Large difference are also detected in the case of an HB with a very small fraction of red HB stars and a large fraction of RR\,Lyrae variables ($f_{\rm R}$=0 or $f_{\rm R}$=0.04). We obtain $\Delta L1=\Delta L2 \sim 0.08$ when assuming $f_{\rm V}$=0.50; $\Delta L1=\Delta L2 \sim 0.04$ for $f_{\rm V}$=0.25; and $\Delta L1=\Delta L2 \sim 0.01$ for $f_{\rm V}$=0.1. We obtain similar results for $\Delta L2$, and $\Delta L1 \sim 0$ when the blue HB hosts a very small fraction of stars ($f_{\rm B}$=0 and $f_{\rm B}$=0.04). $\Delta L1$ and $\Delta L2$ get closer to zero for larger values of $f_{\rm B}$ and $f_{\rm R}$.

According to the literature values (Lee et al.\,1994, Harris\,1996, 2010, Gratton et al.\,2010 and references therein), extreme cases of $f_{\rm V}$$\ge$0.25 and $f_{\rm R}$$\le$ 0.1 are not present among the clusters studied in this paper, which suggests that any error related to the RR\,Lyrae phase should be smaller than $\sim$0.03-0.04 mag.

As a further test, for each GC studied in this paper we assumed the corresponding values of $f_{\rm B}$, $f_{\rm V}$, $f_{\rm R}$ (Lee et al.\,1994, and Harris\,1996, 2003 version, and references therein) and simulated 10,000 CMDs. For most GCs we found $\Delta L1<0.02$ and $\Delta L2<0.02$; $\Delta L1$ is greater than 0.03 mag only in a few GCs, namely NGC\,4590, NGC\,7078, and NGC\,5466. $\Delta L2$ exceeds 0.03 mag also in the cases of Pal\,3 and  Rup\,106. Both $\Delta L1$ and $\Delta L2$ never exceed 0.04 mag.
Our tests suggest that the uncertainties on $L1$ and $L2$ measurements due to the random phase of RR\,Lyrae should be negligible for our purposes. 
This conclusion is similar to that of Gratton et al.\,(2010) who showed that RR\,Lyrae should not affect the determination of the median colors of HB stars.



\begin{thebibliography}{}
\expandafter\ifx\csname natexlab\endcsname\relax\def\natexlab#1{#1}\fi

\bibitem[{{Alves-Brito} {et~al.}(2012){Alves-Brito}, {Yong}, {Mel{\'e}ndez},
  {V{\'a}squez}, \& {Karakas}}]{alv12}
{Alves-Brito}, A., {Yong}, D., {Mel{\'e}ndez}, J., {V{\'a}squez}, S., \&
  {Karakas}, A.~I. 2012, \aap, 540, A3

\bibitem[{{Anderson} {et~al.}(2008){Anderson}, {Sarajedini}, {Bedin}, {King},
  {Piotto}, {Reid}, {Siegel}, {Majewski}, {Paust}, {Aparicio}, {Milone},
  {Chaboyer}, \& {Rosenberg}}]{and08}
{Anderson}, J., {Sarajedini}, A., {Bedin}, L.~R., {et~al.} 2008, \aj, 135, 2055

\bibitem[{{Arp} {et~al.}(1952){Arp}, {Baum}, \& {Sandage}}]{Arp1952}
{Arp}, H.~C., {Baum}, W.~A., \& {Sandage}, A.~R. 1952, \aj, 57, 4

\bibitem[{{Bellini} {et~al.}(2010){Bellini}, {Bedin}, {Piotto}, {Milone},
  {Marino}, \& {Villanova}}]{bel10}
{Bellini}, A., {Bedin}, L.~R., {Piotto}, G., {et~al.} 2010, \aj, 140, 631

\bibitem[{{Bellini} {et~al.}(2013){Bellini}, {Piotto}, {Milone}, {King},
  {Renzini}, {Cassisi}, {Anderson}, {Bedin}, {Nardiello}, {Pietrinferni}, \&
  {Sarajedini}}]{bel13}
{Bellini}, A., {Piotto}, G., {Milone}, A.~P., {et~al.} 2013, \apj, 765, 32

\bibitem[{{Caloi} \& {D'Antona}(2008)}]{cd08}
{Caloi}, V. \& {D'Antona}, F. 2008, \apj, 673, 847

\bibitem[{{Carretta}(2006)}]{car06}
{Carretta}, E. 2006, \aj, 131, 1766

\bibitem[{Carretta} {et~al.}(2007)]{car07} 
{Carretta}, E., {Recio-Blanco}, A., {Gratton}, R.~G., {Piotto}, G., \& {Bragaglia}, A.\ 2007, \apjl, 671, L125 

\bibitem[{Carretta} {et~al.}(2009)]{car09} {Carretta}, E., {Bragaglia}, A., {Gratton}, R., {D'Orazi}, V., \& {Lucatello}, S.\ 2009, \aap, 508, 695 

\bibitem[{{Carretta} {et~al.}(2010{\natexlab{a}}){Carretta}, {Bragaglia},
  {Gratton}, {Lucatello}, {Bellazzini}, {Catanzaro}, {Leone}, {Momany},
  {Piotto}, \& {D'Orazi}}]{car10b}
{Carretta}, E., {Bragaglia}, A., {Gratton}, R.~G., {et~al.} 2010{\natexlab{a}},
  \aap, 520, A95

\bibitem[{{Carretta} {et~al.}(2010{\natexlab{b}}){Carretta}, {Bragaglia},
  {Gratton}, {Recio-Blanco}, {Lucatello}, {D'Orazi}, \& {Cassisi}}]{car10a}
{Carretta}, E., {Bragaglia}, A., {Gratton}, R.~G., {et~al.} 2010{\natexlab{b}},
  \aap, 516, A55

\bibitem[{{Cassisi} {et~al.}(2008){Cassisi}, {Salaris}, {Pietrinferni}, {Piotto}, {Milone}, {Bedin}, \& {Anderson}}]{cas08}
{Cassisi}, S., {Salaris}, M., {Pietrinferni}, A., {et~al.} 2008, \apjl, 672,
  L115

\bibitem[{{Cassisi} {et~al.}(2012){Cassisi}, {Salaris}, {Pietrinferni}}]{cas12}
 {Cassisi}, S., {Salaris}, M., {Pietrinferni}, A. 2012, arXiv:1212.6839

\bibitem[{{Catelan}(2000)}]{cat00}
{Catelan}, M. 2000, \apj, 531, 826

\bibitem[{{Catelan}(2009)}]{c09}
{Catelan}, M. 2009, \apss, 320, 261

\bibitem[{{Catelan} \& {de Freitas Pacheco}(1993)}]{cfp93}
{Catelan}, M. \& {de Freitas Pacheco}, J.~A. 1993, \aj, 106, 1858

\bibitem[{{Dalessandro} {et~al.} (2013)}]{dal13} 
{Dalessandro}, E., {Salaris}, M., {Ferraro}, F.~R., {Mucciarelli}, A., \& {Cassisi}, S. 2013, \mnras, 430, 459 

\bibitem[{D'Antona} \& {Caloi}(2008)]{da08} {D'Antona}, F., \& {Caloi}, V.\ 2008, \mnras, 390, 693 

\bibitem[{{D'Antona} {et~al.}(2005){D'Antona}, {Bellazzini}, {Caloi}, {Pecci},
  {Galleti}, \& {Rood}}]{da05}
{D'Antona}, F., {Bellazzini}, M., {Caloi}, V., {et~al.} 2005, \apj, 631, 868

\bibitem[{{D'Antona} {et~al.}(2002){D'Antona}, {Caloi}, {Montalb{\'a}n},
  {Ventura}, \& {Gratton}}]{da02}
{D'Antona}, F., {Caloi}, V., {Montalb{\'a}n}, J., {Ventura}, P., \& {Gratton},
  R. 2002, \aap, 395, 69

\bibitem[{{De Angeli} {et~al.}(2005){De Angeli}, {Piotto}, {Cassisi}, {Busso},
  {Recio-Blanco}, {Salaris}, {Aparicio}, \& {Rosenberg}}]{de05}
{De Angeli}, F., {Piotto}, G., {Cassisi}, S., {et~al.} 2005, \aj, 130, 116

\bibitem[{di Criscienzo} {et~al.}(2010)]{dic10} 
{di Criscienzo}, M., {Ventura}, P., {D'Antona}, F., {Milone}, A., \& {Piotto}, G.\ 2010, \mnras, 408, 999 

\bibitem[{{di Criscienzo} {et~al.}(2011){di Criscienzo}, {D'Antona}, {Milone},
  {Ventura}, {Caloi}, {Carini}, {D'Ercole}, {Vesperini}, \& {Piotto}}]{dic11}
{di Criscienzo}, M., {D'Antona}, F., {Milone}, A.~P., {et~al.} 2011, \mnras,
  414, 3381

\bibitem[{{Djorgovski} \& {Meylan}(1994)}]{dm94}
{Djorgovski}, S. \& {Meylan}, G. 1994, \aj, 108, 1292

\bibitem[{{Dotter} {et~al.}(2011){Dotter}, {Sarajedini}, \& {Anderson}}]{d11}
{Dotter}, A., {Sarajedini}, A., \& {Anderson}, J. 2011, \apj, 738, 74

\bibitem[{{Dotter} {et~al.}(2010){Dotter}, {Sarajedini}, {Anderson},
  {Aparicio}, {Bedin}, {Chaboyer}, {Majewski}, {Mar{\'{\i}}n-Franch}, {Milone},
  {Paust}, {Piotto}, {Reid}, {Rosenberg}, \& {Siegel}}]{d10}
{Dotter}, A., {Sarajedini}, A., {Anderson}, J., {et~al.} 2010, \apj, 708, 698

\bibitem[{Dotter} {et~al.}(2008)]{d08} 
{Dotter}, A., {Chaboyer}, B., {Jevremovi{\'c}}, D., {et~al.} 2008, \apjs, 178, 89 

\bibitem[{{Ferraro} {et~al.}(1998){Ferraro}, {Paltrinieri}, {Pecci}, {Rood}, \&
  {Dorman}}]{f98}
{Ferraro}, F.~R., {Paltrinieri}, B., {Pecci}, F.~F., {Rood}, R.~T., \&
  {Dorman}, B. 1998, \apj, 500, 311

\bibitem[{{Freeman} \& {Norris}(1981)}]{fn81}
{Freeman}, K.~C. \& {Norris}, J. 1981, \araa, 19, 319

\bibitem[{{Fusi Pecci} {et~al.}(1993){Fusi Pecci}, {Ferraro}, {Bellazzini},
  {Djorgovski}, {Piotto}, \& {Buonanno}}]{fp93}
{Fusi Pecci}, F., {Ferraro}, F.~R., {Bellazzini}, M., {et~al.} 1993, \aj, 105,
  1145

\bibitem[{{Fusi-Pecci} \& {Renzini}(1978)}]{fpr78}
{Fusi Pecci}, F. \& {Renzini}, A. 1978, in IAU Symposium, Vol.~80, The HR
  Diagram - The 100th Anniversary of Henry Norris Russell, ed. A.~G.~D.
  {Philip} \& D.~S. {Hayes}, 225

\bibitem[{{Gratton} {et~al.}(2010){Gratton}, {Carretta},
  {Bragaglia}, {Lucatello}, \& {D'Orazi}}]{g10}
{Gratton}, R.~G., {Carretta}, E., {Bragaglia}, A., {Lucatello}, S., \&
  {D'Orazi}, V. 2010, \aap, 517, A81

\bibitem[{{Gratton} {et~al.}(2011){Gratton}, {Lucatello}, {Carretta},
  {Bragaglia}, {D'Orazi D'Orazi}, \& {Momany}}]{gra11}
{Gratton}, R.~G., {Lucatello}, S., {Carretta}, E., {et~al.} 2011, \aap, 534,
  A123

\bibitem[{{Gratton} {et~al.}(2012){Gratton}, {Lucatello}, {Carretta},
  {Bragaglia}, {D'Orazi}, {Al Momany}, {Sollima}, {Salaris}, \&
  {Cassisi}}]{gra12}
{Gratton}, R.~G., {Lucatello}, S., {Carretta}, E., {et~al.} 2012, \aap, 539,
  A19

\bibitem[{{Gratton} {et~al.}(2013){Gratton}, {Lucatello}, {Sollima},
  {Carretta}, {Bragaglia}, {Momany}, {D'Orazi}, {Cassisi}, {Pietrinferni}, \&
  {Salaris}}]{gra13}
{Gratton}, R.~G., {Lucatello}, S., {Sollima}, A., {et~al.} 2013, \aap, 549, A41

\bibitem[{{Harris}(1996)}]{har96}
{Harris}, W.~E. 1996, \aj, 112, 1487

\bibitem[Iben(1968)]{i68} {Iben}, I.\ 1968, \nat, 220, 143 

\bibitem[{{King}(1962)}]{kin62}
{King}, I. 1962, \aj, 67, 471

\bibitem[{{King} {et~al.}(2012){King}, {Bedin}, {Cassisi}, {Milone}, {Bellini},
  {Piotto}, {Anderson}, {Pietrinferni}, \& {Cordier}}]{kin12}
{King}, I.~R., {Bedin}, L.~R., {Cassisi}, S., {et~al.} 2012, \aj, 144, 5

\bibitem[{Leaman} {et~al.}(2013)]{lea13} 
{Leaman}, R., {VandenBerg}, D.~A., \& {Mendel}, J.~T.\ 2013, arXiv:1309.0822 

\bibitem[{Lee} {et~al.}(1994)]{ldz94}
{Lee}, Y.-W., {Demarque}, P., \& {Zinn}, R. 1994, \apj, 423, 248

\bibitem[{{Lind} {et~al.}(2011){Lind}, {Charbonnel}, {Decressin}, {Primas},
  {Grundahl}, \& {Asplund}}]{lin11}
{Lind}, K., {Charbonnel}, C., {Decressin}, T., {et~al.} 2011, \aap, 527, A148

\bibitem[{{Lovisi} {et~al.}(2012){Lovisi}, {Mucciarelli}, {Lanzoni}, {Ferraro},
  {Gratton}, {Dalessandro}, \& {Contreras Ramos}}]{lov12}
{Lovisi}, L., {Mucciarelli}, A., {Lanzoni}, B., {et~al.} 2012, \apj, 754, 91

\bibitem[{{Mar{\'{\i}}n-Franch} {et~al.}(2009){Mar{\'{\i}}n-Franch},
  {Aparicio}, {Piotto}, {Rosenberg}, {Chaboyer}, {Sarajedini}, {Siegel},
  {Anderson}, {Bedin}, {Dotter}, {Hempel}, {King}, {Majewski}, {Milone},
  {Paust}, \& {Reid}}]{mf09}
{Mar{\'{\i}}n-Franch}, A., {Aparicio}, A., {Piotto}, G., {et~al.} 2009, \apj,
  694, 1498

\bibitem[{{Marino} {et~al.}(2013a){Marino}, {Milone}, \& {Lind}}]{mar13a}
{Marino}, A.~F., {Milone}, A.~P., \& {Lind}, K. 2013, \apj, 768, 27

\bibitem[{Marino} {et~al.}(2013b)]{mar13b} {Marino}, A.~F., {Milone}, 
A.~P., {Przybilla}, N., et al.\ 2013, \mnras, 2693 

\bibitem[{{Marino} {et~al.}(2009){Marino}, {Milone}, {Piotto}, {Villanova},
  {Bedin}, {Bellini}, \& {Renzini}}]{mar09}
{Marino}, A.~F., {Milone}, A.~P., {Piotto}, G., {et~al.} 2009, \aap, 505, 1099

\bibitem[{{Marino} {et~al.}(2012){Marino}, {Milone}, {Sneden}, {Bergemann},
  {Kraft}, {Wallerstein}, {Cassisi}, {Aparicio}, {Asplund}, {Bedin}, {Hilker},
  {Lind}, {Momany}, {Piotto}, {Roederer}, {Stetson}, \& {Zoccali}}]{mar12}
{Marino}, A.~F., {Milone}, A.~P., {Sneden}, C., {et~al.} 2012, \aap, 541, A15

\bibitem[{{Marino} {et~al.}(2011{\natexlab{a}}){Marino}, {Sneden}, {Kraft},
  {Wallerstein}, {Norris}, {da Costa}, {Milone}, {Ivans}, {Gonzalez},
  {Fulbright}, {Hilker}, {Piotto}, {Zoccali}, \& {Stetson}}]{mar11b}
{Marino}, A.~F., {Sneden}, C., {Kraft}, R.~P., {et~al.} 2011{\natexlab{a}},
  \aap, 532, A8

\bibitem[{{Marino} {et~al.}(2011{\natexlab{b}}){Marino}, {Villanova}, {Milone},
  {Piotto}, {Lind}, {Geisler}, \& {Stetson}}]{mar11a}
{Marino}, A.~F., {Villanova}, S., {Milone}, A.~P., {et~al.} 2011{\natexlab{b}},
  \apjl, 730, L16

\bibitem[{{Marino} {et~al.}(2008){Marino}, {Villanova}, {Piotto}, {Milone},
  {Momany}, {Bedin}, \& {Medling}}]{mar08}
{Marino}, A.~F., {Villanova}, S., {Piotto}, G., {et~al.} 2008, \aap, 490, 625

\bibitem[{{Mengel} \& {Gross}(1976)}]{mg76}
{Mengel}, J.~G. \& {Gross}, P.~G. 1976, \apss, 41, 407

\bibitem[{{Milone} {et~al.}(2008){Milone}, {Bedin}, {Piotto}, {Anderson},
  {King}, {Sarajedini}, {Dotter}, {Chaboyer}, {Mar{\'{\i}}n-Franch},
  {Majewski}, {Aparicio}, {Hempel}, {Paust}, {Reid}, {Rosenberg}, \&
  {Siegel}}]{mil08}
{Milone}, A.~P., {Bedin}, L.~R., {Piotto}, G., {et~al.} 2008, \apj, 673, 241

\bibitem[{{Milone} {et~al.}(2013){Milone}, {Marino}, {Piotto}, {Bedin},
  {Anderson}, {Aparicio}, {Bellini}, {Cassisi}, {D'Antona}, {Grundahl},
  {Monelli}, \& {Yong}}]{mil13}
{Milone}, A.~P., {Marino}, A.~F., {Piotto}, G., {et~al.} 2013, \apj, 767, 120

\bibitem[{{Milone} {et~al.}(2012{\natexlab{a}}){Milone}, {Marino}, {Piotto},
  {Bedin}, {Anderson}, {Aparicio}, {Cassisi}, \& {Rich}}]{mil12d}
{Milone}, A.~P., {Marino}, A.~F., {Piotto}, G., {et~al.} 2012{\natexlab{a}},
  \apj, 745, 27

\bibitem[{{Milone} {et~al.}(2012{\natexlab{b}}){Milone}, {Piotto}, {Bedin},
  {Aparicio}, {Anderson}, {Sarajedini}, {Marino}, {Moretti}, {Davies},
  {Chaboyer}, {Dotter}, {Hempel}, {Mar{\'{\i}}n-Franch}, {Majewski}, {Paust},
  {Reid}, {Rosenberg}, \& {Siegel}}]{mil12b}
{Milone}, A.~P., {Piotto}, G., {Bedin}, L.~R., {et~al.} 2012{\natexlab{b}},
  \aap, 540, A16

\bibitem[{{Milone} {et~al.}(2012{\natexlab{c}}){Milone}, {Piotto}, {Bedin},
  {Cassisi}, {Anderson}, {Marino}, {Pietrinferni}, \& {Aparicio}}]{mil12a}
{Milone}, A.~P., {Piotto}, G., {Bedin}, L.~R., {et~al.} 2012{\natexlab{c}},
  \aap, 537, A77

\bibitem[{{Milone} {et~al.}(2012{\natexlab{d}}){Milone}, {Piotto}, {Bedin},
  {King}, {Anderson}, {Marino}, {Bellini}, {Gratton}, {Renzini}, {Stetson},
  {Cassisi}, {Aparicio}, {Bragaglia}, {Carretta}, {D'Antona}, {Di Criscienzo},
  {Lucatello}, {Monelli}, \& {Pietrinferni}}]{mil12c}
{Milone}, A.~P., {Piotto}, G., {Bedin}, L.~R., {et~al.} 2012{\natexlab{d}},
  \apj, 744, 58

\bibitem[{{Monelli} {et~al.}(2013){Monelli}, {Milone}, {Stetson}, {Marino},
  {Cassisi}, {del Pino Molina}, {Salaris}, {Aparicio}, {Asplund}, {Grundahl},
  {Piotto}, {Weiss}, {Carrera}, {Cebri{\'a}n}, {Murabito}, {Pietrinferni}, \&
  {Sbordone}}]{mon13}
{Monelli}, M., {Milone}, A.~P., {Stetson}, P.~B., {et~al.} 2013, \mnras, 431,
  2126

\bibitem[{{Norris}(1981)}]{n81}
{Norris}, J. 1981, \apj, 248, 177

\bibitem[{{Norris}(1983)}]{n83}
{Norris}, J. 1983, \apj, 272, 245

\bibitem[{{Norris}(1987)}]{n87}
{Norris}, J. 1987, \apjl, 313, L65

\bibitem[{{Norris} {et~al.}(1981){Norris}, {Cottrell}, {Freeman}, \& {Da
  Costa}}]{n81b}
{Norris}, J., {Cottrell}, P.~L., {Freeman}, K.~C., \& {Da Costa}, G.~S. 1981,
  \apj, 244, 205

\bibitem[{{Peterson}(1982)}]{pet82}
{Peterson}, R.~C. 1982, \apj, 258, 499

\bibitem[{{Piotto} {et~al.}(2002){Piotto}, {King}, {Djorgovski}, {Sosin},
  {Zoccali}, {Saviane}, {De Angeli}, {Riello}, {Recio-Blanco}, {Rich},
  {Meylan}, \& {Renzini}}]{pio02}
{Piotto}, G., {King}, I.~R., {Djorgovski}, S.~G., {et~al.} 2002, \aap, 391, 945

\bibitem[{{Piotto} {et~al.}(2007)}]{pio07} {Piotto}, G., {Bedin}, 
L.~R., {Anderson}, J., {et~al}.\ 2007, \apjl, 661, L53 

\bibitem[{{Piotto} {et~al.}(2012){Piotto}, {Milone}, {Anderson}, {Bedin},
  {Bellini}, {Cassisi}, {Marino}, {Aparicio}, \& {Nascimbeni}}]{pio12}
{Piotto}, G., {Milone}, A.~P., {Anderson}, J., {et~al.} 2012, \apj, 760, 39

\bibitem[{Piotto} {et~al.}(2013)]{pio13} {Piotto}, G., {Milone}, 
A.~P., {Marino}, A.~F., et al.\ 2013, arXiv:1306.5795 

\bibitem[{{Recio-Blanco} {et~al.}(2006){Recio-Blanco}, {Aparicio}, {Piotto},
  {de Angeli}, \& {Djorgovski}}]{rb06}
{Recio-Blanco}, A., {Aparicio}, A., {Piotto}, G., {de Angeli}, F., \&
  {Djorgovski}, S.~G. 2006, \aap, 452, 875

\bibitem[{{Rood} \& {Seitzer}(1981)}]{rs81}
{Rood}, R.~T. \& {Seitzer}, P.~O. 1981, in IAU Colloq. 68: Astrophysical
  Parameters for Globular Clusters, ed. A.~G.~D. {Philip} \& D.~S. {Hayes}, 369

\bibitem[{{Rosenberg} {et~al.}(1999){Rosenberg}, {Saviane}, {Piotto}, \&
  {Aparicio}}]{ros99}
{Rosenberg}, A., {Saviane}, I., {Piotto}, G., \& {Aparicio}, A. 1999, \aj, 118,
  2306
\bibitem[Salaris et al.(2004)]{sal04} {Salaris}, M., {Riello}, M., {Cassisi}, S., \& {Piotto}, G.\ 2004, \aap, 420, 911 

\bibitem[{{Sandage} \& {Wallerstein}(1960)}]{sw60}
{Sandage}, A. \& {Wallerstein}, G. 1960, \apj, 131, 598

\bibitem[{{Sandage} \& {Wildey}(1967)}]{sw67}
{Sandage}, A. \& {Wildey}, R. 1967, \apj, 150, 469

\bibitem[{{Sarajedini} {et~al.}(2007){Sarajedini}, {Bedin}, {Chaboyer},
  {Dotter}, {Siegel}, {Anderson}, {Aparicio}, {King}, {Majewski},
  {Mar{\'{\i}}n-Franch}, {Piotto}, {Reid}, \& {Rosenberg}}]{sar07}
{Sarajedini}, A., {Bedin}, L.~R., {Chaboyer}, B., {et~al.} 2007, \aj, 133, 1658

\bibitem[{{Searle} \& {Zinn}(1978)}]{sz78}
{Searle}, L. \& {Zinn}, R. 1978, \apj, 225, 357

\bibitem[{{Shetrone} \& {Keane}(2000)}]{she00}
{Shetrone}, M.~D. \& {Keane}, M.~J. 2000, \aj, 119, 840

\bibitem[{{Siess} \& {Livio}(1999)}]{sl99}
{Siess}, L. \& {Livio}, M. 1999, \mnras, 308, 1133

\bibitem[{{Smith} \& {Mateo}(1990)}]{smi90}
{Smith}, G.~H. \& {Mateo}, M. 1990, \apj, 353, 533

\bibitem[{{Smith} \& {Norris}(1993)}]{sn93}
{Smith}, G.~H. \& {Norris}, J.~E. 1993, \aj, 105, 173

\bibitem[{{Sneden} {et~al.}(2004){Sneden}, {Kraft}, {Guhathakurta}, {Peterson},
  \& {Fulbright}}]{sne04}
{Sneden}, C., {Kraft}, R.~P., {Guhathakurta}, P., {Peterson}, R.~C., \&
  {Fulbright}, J.~P. 2004, \aj, 127, 2162

\bibitem[{Sirianni} {et~al.}(2005)]{si05} 
{Sirianni}, M., {Jee}, M.~J., {Ben{\'{\i}}tez}, N., {et~al.} 2005, \pasp, 117, 1049 

\bibitem[{{Soker}(1998)}]{s98}
{Soker}, N. 1998, \aj, 116, 1308

\bibitem[{{Stetson} {et~al.}(1999)}]{ste99} 
{Stetson}, P.~B., {Bolte}, M., {Harris}, W.~E., {et~al.} 1999, \aj, 117, 247 

\bibitem[{{van den Bergh}(1965)}]{vdb65}
{van den Bergh}, S. 1965, \jrasc, 59, 151

\bibitem[{{van den Bergh}(1967)}]{vdb67}
{van den Bergh}, S. 1967, \aj, 72, 324

\bibitem[{{VandenBerg {et~al.}}(2013)}]{van13}
{Vandenberg}, D.~A., {Brogaard}, K., {Leaman}, R., \& {Casagrande}, L., 2013, arXiv:1308.2257 

\bibitem[{Ventura} \& {D'Antona}(2005)]{vd05} {Ventura}, P., \& {D'Antona}, F.\ 2005, \apjl, 635, L149 

\bibitem[{{Ventura} {et~al.}(2009){Ventura}, {Caloi}, {D'Antona}, {Ferguson},
  {Milone}, \& {Piotto}}]{ven09}
{Ventura}, P., {Caloi}, V., {D'Antona}, F., {et~al.} 2009, \mnras, 399, 934

\bibitem[{{Villanova} {et~al.}(2010){Villanova}, {Geisler}, \&
  {Piotto}}]{vil10}
{Villanova}, S., {Geisler}, D., \& {Piotto}, G. 2010, \apjl, 722, L18

\bibitem[{{Villanova} {et~al.}(2009){Villanova}, {Piotto}, \&
  {Gratton}}]{vil09}
{Villanova}, S., {Piotto}, G., \& {Gratton}, R.~G. 2009, \aap, 499, 755

\bibitem[{{Yong} {et~al.}(2009){Yong}, {Grundahl}, {D'Antona}, {Karakas},
  {Lattanzio}, \& {Norris}}]{yon09}
{Yong}, D., {Grundahl}, F., {D'Antona}, F., {et~al.} 2009, \apjl, 695, L62

\bibitem[{{Yong} {et~al.}(2008){Yong}, {Grundahl}, {Johnson}, \&  {Asplund}}]{yon08}
{Yong}, D., {Grundahl}, F., {Johnson}, J.~A., \& {Asplund}, M. 2008, \apj, 684,
  1159

\end{thebibliography}

\clearpage

\begin{table}[!htp]
\center
\scriptsize {
\begin{tabular}{lcccccc}
\hline
 ID & $P_{\rm A}$ & $P_{\rm B}$ & $P_{\rm C}$ & $L1$ & $L2$ & group \\
\hline
 AM\,1&   0.558$\pm$0.003 & 0.627$\pm$0.003 & 0.720$\pm$0.006 & 0.093$\pm$0.007 & 0.069$\pm$0.004 & G2 \\
ARP\,2    &   0.105$\pm$0.013 & 0.289$\pm$0.020 & 0.780$\pm$0.003 & 0.491$\pm$0.021 & 0.184$\pm$0.021 & G3 \\
 ERIDANUS  &   0.584$\pm$0.011 & 0.642$\pm$0.005 & 0.755$\pm$0.011 & 0.111$\pm$0.009 & 0.058$\pm$0.012 & G2 \\
IC\,4499  &   0.307$\pm$0.031 & 0.815$\pm$0.024 & 0.928$\pm$0.008 & 0.113$\pm$0.026 & 0.508$\pm$0.041 & G2 \\
LYNGA\,7  &   1.387$\pm$0.008 & 1.480$\pm$0.003 & 1.534$\pm$0.004 & 0.055$\pm$0.050 & 0.093$\pm$0.009 & G1 \\
\hline                                                                         
\end{tabular}                 
 }
\caption{Horizontal Branch parameters.} 
\end{table}

\begin{table}[!htp]
\center
\scriptsize {
\begin{tabular}{lcc}
\hline
\hline
 ID & $\Delta$Y & Reference \\
\hline
NGC\,104 (47\,Tuc) & $\sim$0.03 & di Criscienzo et al.\,(2010), Milone et al.\,(2012d)\\
NGC\,288  & 0.013$\pm$0.001 &  Piotto et al.\ (2013) \\ 
NGC\,2419 & $\sim$0.17 & di Criscienzo et al.\,(2011) \\
NGC\,2808 & $\sim$0.14 & Milone et al.\,(2012c) \\ 
NGC\,5139 ($\omega$ Cen) & 0.14$\pm$0.01 & King et al.\,(2012)\\ 
NGC\,6397 & $\sim$0.01 & Milone et al.\,(2012a)$^{*}$\\
NGC\,6441 & $\sim$0.07 & Bellini et al.\,(2013)\\
NGC\,6752 & 0.035$\pm$0.012 & Milone et al.\,(2013) \\
NGC\,7078 & 0.053$\pm$0.015 & Milone et al.\ (2014, in preparation) \\
\hline
\hline
\end{tabular}\\
$^{*}$Accurate analysis of the MS width by di Criscienzo et al.\,(2010) previously showed that any helium variation in NGC\,6397 must be smaller than $\Delta Y \sim$0.02.
}
\label{tab:el}
\caption{Literature estimate of the maximum helium difference between stellar populations based on the analysis of multiple MSs or multiple RGBs.}
\end{table}

\clearpage 
\clearpage

\begin{table}
\label{tab:r}
\scriptsize {
\begin{tabular}{|c|cccc|cccc|}
\hline
              &        &  $L1$   &        &          &         & $L2$   &         &          \\
 Parameter    & $G1$ & $G2$ & $G3$ & $G2+G3$ & $G1$ & $G2$ & $G3$ & $G2+G3$ \\
\hline
\hline
[Fe/H] & $r=-$0.13  &   $r=-$0.70 & $r=$0.42 & $r=-$0.47     & $r=-$0.08  &   $r=$0.19 & $r=$0.57 & $r=$0.24  \\
   & $N$=15   & $N$=38  & $N$=21 & $N$=59 & $N$=15   & $N$=38  & $N$=21 & $N$=59 \\
       & $\sigma_{\rm r}$=0.26  &   $\sigma_{\rm r}$=0.08 & $\sigma_{\rm r}$=0.18 & $\sigma_{\rm r}$=0.11     & $\sigma_{\rm r}$=0.31  &   $\sigma_{\rm r}$=0.16 & $\sigma_{\rm r}$=0.13 & $\sigma_{\rm r}$=0.13  \\
\hline
$M_{\rm V}$ & $r=$0.09  &   $r=-$0.23 & $r=-$0.08 & $r=$0.09     & $r=-$0.57  &   $r=-$0.89 & $r=-$0.71 & $r=-$0.80  \\
   & $N$=15   & $N$=38  & $N$=21 & $N$=59 & $N$=15   & $N$=38  & $N$=21 & $N$=59 \\
       & $\sigma_{\rm r}$=0.13  &   $\sigma_{\rm r}$=0.16 & $\sigma_{\rm r}$=0.24 & $\sigma_{\rm r}$=0.13     & $\sigma_{\rm r}$=0.23  &   $\sigma_{\rm r}$=0.05 & $\sigma_{\rm r}$=0.13 & $\sigma_{\rm r}$=0.06  \\
\hline
$AGE_{\rm D10}$ & $r=$0.35  &   $r=$0.76 & $r=-$0.22 & $r=$0.72     & $r=$0.23  &   $r=$0.28 & $r=-$0.07 & $r=-$0.06  \\
    & $N$=15  & $N$=37 & $N$=21 & $N$=58 & $N$=15  & $N$=37 & $N$=21 & $N$=58 \\
       & $\sigma_{\rm r}$=0.20  &   $\sigma_{\rm r}$=0.08 & $\sigma_{\rm r}$=0.20 & $\sigma_{\rm r}$=0.08     & $\sigma_{\rm r}$=0.29  &   $\sigma_{\rm r}$=0.18 & $\sigma_{\rm r}$=0.24 & $\sigma_{\rm r}$=0.16  \\
\hline
$AGE_{\rm MF09}$& $r=$0.38  &   $r=$0.47 & $r=-$0.05 & $r=$0.38     & $r=$0.30  &   $r=-$0.29 & $r=$0.17 & $r=-$0.30  \\
    & $N$=15  & $N$=26 & $N$=20 & $N$=46 & $N$=15  & $N$=26 & $N$=20 & $N$=46  \\
       & $\sigma_{\rm r}$=0.26  &   $\sigma_{\rm r}$=0.18 & $\sigma_{\rm r}$=0.20 & $\sigma_{\rm r}$=0.15     & $\sigma_{\rm r}$=0.30  &   $\sigma_{\rm r}$=0.16 & $\sigma_{\rm r}$=0.22 & $\sigma_{\rm r}$=0.14  \\
\hline
$AGE_{\rm DA05}$& $r=$0.60  &   $r=$0.72 & $r=-$0.32 & $r=$0.57     & $r=-$0.60  &   $r=-$0.50 & $r=$0.03 & $r=-$0.49  \\
    & $N$=5   & $N$=25 & $N$=11 & $N$=36 & $N$=5   & $N$=25 & $N$=11 & $N$=36 \\
       & $\sigma_{\rm r}$=0.37  &   $\sigma_{\rm r}$=0.13 & $\sigma_{\rm r}$=0.36 & $\sigma_{\rm r}$=0.13     & $\sigma_{\rm r}$=0.40  &   $\sigma_{\rm r}$=0.14 & $\sigma_{\rm r}$=0.36 & $\sigma_{\rm r}$=0.13  \\
\hline
 AGE$_{\rm V13}$ & $r=$0.62  &   $r=$0.73 & $r=-$0.21     & $r=$0.68  &   $r=$0.62 & $r=$0.01 & $r=-$0.22 & $r=-$0.34  \\
    & $N$=12 & $N$=31  & $N$=18 & $N$=49 & $N$=12    & $N$=31 & $N$=18 & $N$=49 \\
       & $\sigma_{\rm r}$=0.20  &   $\sigma_{\rm r}$=0.09 & $\sigma_{\rm r}$=0.25 & $\sigma_{\rm r}$=0.10     & $\sigma_{\rm r}$=0.21  &   $\sigma_{\rm r}$=0.20 & $\sigma_{\rm r}$=0.25 & $\sigma_{\rm r}$=0.16  \\
\hline
$AGE_{\rm D10}^{*}$ & $r=$0.35  &   $r=$0.76 & $r=-$0.22     & $r=$0.74  &   $r=$0.39 & $r=$0.30 & $r=-$0.08 & $r=-$0.08  \\
    & $N$=14 & $N$=34  & $N$=21 & $N$=55 & $N$=14    & $N$=34 & $N$=21 & $N$=55 \\
       & $\sigma_{\rm r}$=0.20  &   $\sigma_{\rm r}$=0.08 & $\sigma_{\rm r}$=0.20 & $\sigma_{\rm r}$=0.07     & $\sigma_{\rm r}$=0.26  &   $\sigma_{\rm r}$=0.19 & $\sigma_{\rm r}$=0.24 & $\sigma_{\rm r}$=0.17  \\
\hline
$AGE_{\rm MF09}^{*}$ & $r=$0.41  &   $r=$0.40  & $r=-$0.05     & $r=$0.33  &   $r=$0.31 & $r=-$0.29 & $r=$0.17 & $r=-$0.30  \\
    & $N$=14  &  $N$=23 & $N$=20 & $N$=43 & $N$=14  &  $N$=23 & $N$=20 & $N$=43   \\
       & $\sigma_{\rm r}$=  &   $\sigma_{\rm r}$=0.21 & $\sigma_{\rm r}$=0.20 & $\sigma_{\rm r}$=0.16     & $\sigma_{\rm r}$=0.30  &   $\sigma_{\rm r}$=0.16 & $\sigma_{\rm r}$=0.22 & $\sigma_{\rm r}$=0.14  \\
\hline
$AGE_{\rm DA05}^{*}$ & $r=$0.60  &   $r=$0.71 & $r=-$0.32 & $r=$0.53     & $r=-$0.60  &   $r=-$0.53 & $r=$0.03 & $r=-$0.47  \\
    & $N$=5  & $N$=23 &  $N$=11 & $N$=34 & $N$=5  & $N$=23 &  $N$=11 & $N$=34    \\
       & $\sigma_{\rm r}$=0.37  &   $\sigma_{\rm r}$=0.14 & $\sigma_{\rm r}$=0.36 & $\sigma_{\rm r}$=0.15     & $\sigma_{\rm r}$=0.40  &   $\sigma_{\rm r}$=0.14 & $\sigma_{\rm r}$=0.36 & $\sigma_{\rm r}$=0.13  \\
\hline
 AGE$_{\rm V13}^{*}$ & $r=$0.62  &   $r=$0.70 & $r=-$0.21     & $r=$0.65  &   $r=$0.62 & $r=$0.07 & $r=-$0.22 & $r=-$0.32  \\
    & $N$=12 & $N$=28  & $N$=18 & $N$=46 & $N$=12    & $N$=28 & $N$=18 & $N$=46 \\
       & $\sigma_{\rm r}$=0.20  &   $\sigma_{\rm r}$=0.10 & $\sigma_{\rm r}$=0.25 & $\sigma_{\rm r}$=0.10     & $\sigma_{\rm r}$=0.21  &   $\sigma_{\rm r}$=0.22 & $\sigma_{\rm r}$=0.25 & $\sigma_{\rm r}$=0.16  \\
\hline
log($\Delta$Y) & ---  &   $r=-$0.30 & --- & $r=-$0.71 & ---  &   $r=$0.70 & --- & $r=$0.89       \\
    & $N$=2  & $N$=5 & $N$=2 &  $N$=7   & $N$=2 & $N$=5 & $N$=2 & $N$=7 \\
               & ---  &   $\sigma_{\rm r}$=0.54 & --- & $\sigma_{\rm r}$=0.31 &  ---  &   $\sigma_{\rm r}$=0.37 & --- & $\sigma_{\rm r}$=0.17 \\
\hline
$\sigma_{\rm V}$ & $r=-$0.54 & $r=-$0.07 & $r=$  0.10 & $r=-$0.02   & $r=$0.66 & $r=$0.79 & $r=$0.40 & $r=$0.46 \\
 & $N$=6 & $N$=7 & $N$=13 & $N$=38   & $N$=6 & $N$=7 & $N$=13 & $N$=38   \\
       & $\sigma_{\rm r}$=0.38  &   $\sigma_{\rm r}$=0.22 & $\sigma_{\rm r}$=0.30 & $\sigma_{\rm r}$=0.15     & $\sigma_{\rm r}$=0.29  &   $\sigma_{\rm r}$=0.12 & $\sigma_{\rm r}$=0.28 & $\sigma_{\rm r}$=0.14  \\
\hline
c            & $r=$0.13 & $r=$0.08 & $r=$0.26 & $r=-$0.10    & $r=$  0.07 & $r=$  0.54 & $r=$  0.52 & $r=$  0.55 \\
            & $N$=14 & $N$=34 & $N$=16 & $N$=50   & $N$=14 & $N$=34 & $N$=16 & $N$=50 \\
       & $\sigma_{\rm r}$=0.25  &   $\sigma_{\rm r}$=0.17 & $\sigma_{\rm r}$=0.23 & $\sigma_{\rm r}$=0.14     & $\sigma_{\rm r}$=0.31  &   $\sigma_{\rm r}$=0.14 & $\sigma_{\rm r}$=0.23 & $\sigma_{\rm r}$=0.10  \\
\hline
$\mu_{\rm V}$ & $r=$$-$0.01  &   $r=$$-$0.07 & $r=$$-$0.23 & $r=$$-$0.07     & $r=$$-$0.65  &   $r=$$-$0.76 & $r=$$-$0.68 & $r=$$-$0.57  \\
    & $N$=14  &  $N$=37 & $N$=21 & $N$=58    & $N$=14  &  $N$=37 & $N$=21 & $N$=58 \\
       & $\sigma_{\rm r}$=0.28  &   $\sigma_{\rm r}$=0.18 & $\sigma_{\rm r}$=0.23 & $\sigma_{\rm r}$=0.14   & $\sigma_{\rm r}$=0.21  &   $\sigma_{\rm r}$=0.10 & $\sigma_{\rm r}$=0.15 & $\sigma_{\rm r}$=0.11  \\
\hline
$\rho$ & $r=-$0.09  &   $r=$0.05 & $r=$0.29 & $r=$0.14     & $r=$0.69  &   $r=$0.69 & $r=$0.56 & $r=$0.47  \\
  &  $N$=14 & $N$=37  & $N$=21 & $N$=58     & $N$=14 & $N$=37  & $N$=21 & $N$=58 \\
       & $\sigma_{\rm r}$=0.26  &   $\sigma_{\rm r}$=0.18 & $\sigma_{\rm r}$=0.21 & $\sigma_{\rm r}$=0.14     & $\sigma_{\rm r}$=0.18  &   $\sigma_{\rm r}$=0.12 & $\sigma_{\rm r}$=0.19 & $\sigma_{\rm r}$=0.13  \\
\hline
$log(\tau_{\rm c})$ & $r=-$0.31  &   $r=$0.09 & $r=-$0.28 & $r=-$0.21     & $r=-$0.15  &   $r=-$0.35 & $r=-0.8$ & $r=-$0.20  \\
    & $N$=14  & $N$=37 & $N$=21 & $N$=58    & $N$=14 & $N$=37 & $N$=21 & $N$=58 \\
       & $\sigma_{\rm r}$=0.21  &   $\sigma_{\rm r}$=0.18 & $\sigma_{\rm r}$=0.20 & $\sigma_{\rm r}$=0.13     & $\sigma_{\rm r}$=0.32  &   $\sigma_{\rm r}$=0.18 & $\sigma_{\rm r}$=0.23 & $\sigma_{\rm r}$=0.15  \\
\hline
$log(\tau_{\rm h})$ & $r=-$0.29  &   $r=$0.19 & $r=-$0.21 & $r=-$0.33     & $r=-$0.28  &   $r=-$0.13 & $r=-$0.23 & $r=$0.08  \\
    &  $N$=15 & $N$=37 & $N$=21 & $N$=58 &  $N$=15 & $N$=37 & $N$=21 & $N$=58  \\
       & $\sigma_{\rm r}$=0.28  &   $\sigma_{\rm r}$=0.18 & $\sigma_{\rm r}$=0.18 & $\sigma_{\rm r}$=0.13     & $\sigma_{\rm r}$=0.27  &   $\sigma_{\rm r}$=0.18 & $\sigma_{\rm r}$=0.24 & $\sigma_{\rm r}$=0.15  \\
\hline
$\epsilon$ & $r=-$0.10  &   $r=$0.45 & $r=$0.14 & $r=$0.07     & $r=$0.05  &   $r=-$0.05 & $r=-$0.16 & $r=$0.09  \\
    & $N$=12  & $N$=30  & $N$=17 & $N$=47 & $N$=12  & $N$=30  & $N$=17 & $N$=47  \\
       & $\sigma_{\rm r}$=0.32  &   $\sigma_{\rm r}$=0.14 & $\sigma_{\rm r}$=0.27 & $\sigma_{\rm r}$=0.15     & $\sigma_{\rm r}$=0.33  &   $\sigma_{\rm r}$=0.20 & $\sigma_{\rm r}$=0.27 & $\sigma_{\rm r}$=0.15  \\
\hline
$log(R_{\rm GC}[kpc])$ & $r=-$0.08  &   $r=-$0.18 & $r=-$0.29 & $r=-$0.45     & $r=-$0.50  &   $r=-$0.44 & $r=-$0.50 & $r=-$0.17  \\
    & $N$=15  & $N$=38 & $N$=21 & $N$=59 & $N$=15  & $N$=38 & $N$=21 & $N$=59 \\
       & $\sigma_{\rm r}$=0.27  &   $\sigma_{\rm r}$=0.16 & $\sigma_{\rm r}$=0.19 & $\sigma_{\rm r}$=0.11     & $\sigma_{\rm r}$=0.22  &   $\sigma_{\rm r}$=0.16 & $\sigma_{\rm r}$=0.19 & $\sigma_{\rm r}$=0.15  \\
\hline
\hline                                                                          
\end{tabular}\\                       
}                                         
$^{*}$ GCs with double SGB excluded from the analysis\\ 

\caption{Spearman's rank correlation coefficients indicating the statistical dependence between $L1$ (columns 2-5), $L2$ (columns 6-9) and several parameters of the host GCs for G1, G2, G3, and G2+G3 GCs. The values of $\sigma_{\rm r}$, which provide an estimate of the robustness of $r$ measurements, and the numbers of analyzed GCs ($N$) are also listed.}
\end{table}

\begin{table}
\label{tab:r2}
\scriptsize {
\begin{tabular}{|c|cccc|cccc|}
\hline
              &        &  $L1$   &        &          &         & $L2$   &         &          \\
 Parameter    & $G1$ & $G2$ & $G3$ & $G2+G3$ & $G1$ & $G2$ & $G3$ & $G2+G3$ \\
\hline
\hline
$S_{\rm RR\,Lyrae}$ & $r=-$0.01  &   $r=-$0.26 & $r=-$0.48 & $r=-$0.66     & $r=-$0.12  &   $r=-$0.68 & $r=-$0.51 & $r=-$0.32  \\
    & $N$=14  & $N$=36 & $N$=20 & $N$=56 & $N$=14  & $N$=36 & $N$=20 & $N$=56  \\
       & $\sigma_{\rm r}$=0.27  &   $\sigma_{\rm r}$=0.17 & $\sigma_{\rm r}$=0.17 & $\sigma_{\rm r}$=0.11     & $\sigma_{\rm r}$=0.22  &   $\sigma_{\rm r}$=0.19 & $\sigma_{\rm r}$=0.24 & $\sigma_{\rm r}$=0.13  \\
\hline
${\it f}_{\rm bin}^{\rm C}$ & $r=$0.19  &   $r=-$0.03 & $r=-$0.22 & $r=-$0.18     & $r=-$0.12  &   $r=-$0.68 & $r=-$0.51 & $r=-$0.32  \\
    & $N$=9  & $N$=16 & $N$=15 & $N$=31 & $N$=9  & $N$=16 & $N$=15 & $N$=31     \\
       & $\sigma_{\rm r}$=0.40  &   $\sigma_{\rm r}$=0.30 & $\sigma_{\rm r}$=0.26 & $\sigma_{\rm r}$=0.18     & $\sigma_{\rm r}$=0.36  &   $\sigma_{\rm r}$=0.15 & $\sigma_{\rm r}$=0.25 & $\sigma_{\rm r}$=0.17  \\
\hline
${\it f}_{\rm bin}^{\rm C-HM}$ & $r=-$0.09  &   $r=-$0.19 & $r=-$0.18 & $r=-$0.15  & $r=-$0.67  &   $r=-$0.47 & $r=-$0.56 & $r=-$0.27  \\
    & $N$=11  & $N$=24 & $N$=16 & $N$=40 & $N$=11  & $N$=24 & $N$=16 & $N$=40    \\
       & $\sigma_{\rm r}$=0.35  &   $\sigma_{\rm r}$=0.19 & $\sigma_{\rm r}$=0.29 & $\sigma_{\rm r}$=0.14     & $\sigma_{\rm r}$=0.23  &   $\sigma_{\rm r}$=0.20 & $\sigma_{\rm r}$=0.23 & $\sigma_{\rm r}$=0.15  \\
\hline
${\it f}_{\rm bin}^{\rm oHM}$ & $r=$0.49  &   $r=-$0.14 & $r=-$0.07 & $r=-$0.02     & $r=-$0.75  &   $r=-$0.45 & $r=-$0.69 & $r=-$0.39  \\
    & $N$=11  & $N$=18 & $N$=17 & $N$=35 & $N$=11  & $N$=18 & $N$=17 & $N$=35 \\
       & $\sigma_{\rm r}$=0.29  &   $\sigma_{\rm r}$=0.24 & $\sigma_{\rm r}$=0.28 & $\sigma_{\rm r}$=0.17     & $\sigma_{\rm r}$=0.18  &   $\sigma_{\rm r}$=0.21 & $\sigma_{\rm r}$=0.19 & $\sigma_{\rm r}$=0.15  \\
\hline
$W_{\rm RGB}$ & ---  &   $r=-$0.45 & $r=-$0.08 & $r=-$0.23     & ---  &   $r=$0.50 & $r=$0.35 & $r=$0.30  \\
   & $N$=3 &  $N$=9 &  $N$=9 & $N$=18 & $N$=3 &  $N$=9 &  $N$=9 & $N$=18 \\
       & ---  &   $\sigma_{\rm r}$=0.31 & $\sigma_{\rm r}$=0.35 & $\sigma_{\rm r}$=0.22     & ---  &   $\sigma_{\rm r}$=0.35 & $\sigma_{\rm r}$=0.37 & $\sigma_{\rm r}$=0.23  \\
\hline
IQR([O/Na]) & ---  &   $r=-$0.13 & $r=$0.00 & $r=-$0.02     & ---  &   $r=$0.82 & $r=$0.90 & $r=$0.41  \\
   & $N$=4 & $N$=13  & $N$=8 & $N$=21 & $N$=4 & $N$=13  & $N$=8 & $N$=21  \\
       & ---  &   $\sigma_{\rm r}$=0.26 & $\sigma_{\rm r}$=0.38 & $\sigma_{\rm r}$=0.21     & ---  &   $\sigma_{\rm r}$=0.13 & $\sigma_{\rm r}$=0.17 & $\sigma_{\rm r}$=0.21  \\
\hline
$R_{\rm CN}$ & ---  &   $r=$0.29 & --- & $r=$0.13     & ---  &   $r=$0.47 & --- & $r=$0.38  \\
   & $N$=2 & $N$=11  & $N$=3 & $N$=14 & $N$=2 & $N$=11  & $N$=3 & $N$=14  \\
       & ---  &   $\sigma_{\rm r}$=0.30 & --- & $\sigma_{\rm r}$=0.27     & ---  &   $\sigma_{\rm r}$=0.24 & --- & $\sigma_{\rm r}$=0.27  \\
\hline
{\it R} & $r=$0.07  &   $r=$0.17 & $r=-$0.07 & $r=$0.11     & $r=$0.26  &   $r=$0.19 & $r=-$0.32 & $r=$0.07  \\
    & $N$=8  & $N$=19 & $N$=7 & $N$=26  & $N$=8  & $N$=19 & $N$=7 & $N$=26   \\
       & $\sigma_{\rm r}$=0.40  &   $\sigma_{\rm r}$=0.26 & $\sigma_{\rm r}$=0.39 & $\sigma_{\rm r}$=0.19     & $\sigma_{\rm r}$=0.35  &   $\sigma_{\rm r}$=0.24 & $\sigma_{\rm r}$=0.42 & $\sigma_{\rm r}$=0.22  \\
\hline
${\it f}_{\rm POPI}$ & ---  &   $r=$0.38 & $r=-$0.32 & $r=-$0.03     & ---  &   $r=-$0.12 & $r=-$0.07 & $r=-$0.02  \\
    & $N$=4  & $N$=8 & $N$=7 & $N$=15 & $N$=4  & $N$=8 & $N$=7 & $N$=15 \\
       & ---  &   $\sigma_{\rm r}$=0.39 & $\sigma_{\rm r}$=0.36 & $\sigma_{\rm r}$=0.29     & ---  &   $\sigma_{\rm r}$=0.38 & $\sigma_{\rm r}$=0.36 & $\sigma_{\rm r}$=0.27  \\
\hline
 Y(R') & $r=-$0.27  &   $r=-$0.16 & $r=$0.12 & $r=$0.07     & $r=$0.15  &   $r=$0.44 & $r=-$0.06 & $r=$0.04  \\
    & $N$=11  & $N$=25 & $N$=17 & $N$=42  & $N$=11  & $N$=25 & $N$=17 & $N$=42   \\
       & $\sigma_{\rm r}$=0.36  &   $\sigma_{\rm r}$=0.23 & $\sigma_{\rm r}$=0.28 & $\sigma_{\rm r}$=0.15     & $\sigma_{\rm r}$=0.34  &   $\sigma_{\rm r}$=0.21 & $\sigma_{\rm r}$=0.26 & $\sigma_{\rm r}$=0.16  \\
\hline
$L2$ & $r=-$0.35  &   $r=$0.05 & $r=$0.10 & $r=-$0.35     & $r=$1.00  &   $r=$1.00 & $r=$1.00 & $r=$1.00  \\
    & $N$=15  & $N$=38 & $N$=21 & $N$=59    & $N$=15 & $N$=38 & $N$=21 & $N$=59 \\
      & $\sigma_{\rm r}$=0.24  &   $\sigma_{\rm r}$=0.17 & $\sigma_{\rm r}$=0.25 & $\sigma_{\rm r}$=0.15     & $\sigma_{\rm r}$=0.00  &   $\sigma_{\rm r}$=0.00 & $\sigma_{\rm r}$=0.00 & $\sigma_{\rm r}$=0.00  \\
 \hline
HBR & $r=$0.43  &   $r=$0.79 & $r=$0.48 & $r=$0.91     & $r=-$0.24  &   $r=$0.44 & $r=-$0.04 & $r=-$0.15  \\
    & $N$=13  & $N$=37 & $N$=21 &  $N$=58 & $N$=13  & $N$=37 & $N$=21 &  $N$=58 \\
      & $\sigma_{\rm r}$=0.28  &   $\sigma_{\rm r}$=0.07 & $\sigma_{\rm r}$=0.16 & $\sigma_{\rm r}$=0.02     & $\sigma_{\rm r}$=0.27  &   $\sigma_{\rm r}$=0.16 & $\sigma_{\rm r}$=0.26 & $\sigma_{\rm r}$=0.17  \\
\hline
$\delta M$ & $r=-$0.48  &   $r=-$0.36 & $r=$0.12 & $r=-$0.19     & $r=$0.39  &   $r=$0.66 & $r=$0.80 & $r=$0.48  \\
    & $N$=12 & $N$=25 & $N$=17 & $N$=42 & $N$=12 & $N$=25 & $N$=17 & $N$=42     \\
       & $\sigma_{\rm r}$=0.24  &   $\sigma_{\rm r}$=0.20 & $\sigma_{\rm r}$=0.29 & $\sigma_{\rm r}$=0.30     & $\sigma_{\rm r}$=0.30  &   $\sigma_{\rm r}$=0.14 & $\sigma_{\rm r}$=0.15 & $\sigma_{\rm r}$=0.13  \\
\hline
$\Delta M_{\rm median}$ & $r=-$0.47  &   $r=-$0.68 & $r=$0.50 & $r=$0.00     & $r=$0.39  &   $r=$0.63 & $r=$0.35 & $r=$0.14  \\
    & $N$=11 & $N$=24  & $N$=17 & $N$=41  & $N$=11 & $N$=24  & $N$=17 & $N$=41    \\
       & $\sigma_{\rm r}$=0.28  &   $\sigma_{\rm r}$=0.10 & $\sigma_{\rm r}$=0.19 & $\sigma_{\rm r}$=0.17     & $\sigma_{\rm r}$=0.30  &   $\sigma_{\rm r}$=0.17 & $\sigma_{\rm r}$=0.22 & $\sigma_{\rm r}$=0.17  \\
\hline
$L_{\rm t}$ & ---  &   $r=-$0.21 & $r=$0.68 & $r=-$0.06     & ---  &   $r=$0.62 & $r=$0.38 & $r=$0.44  \\
    &  $N$=4 & $N$=23 & $N$=11 & $N$=34 &  $N$=4 & $N$=23 & $N$=11 & $N$=34 \\
       & ---  &   $\sigma_{\rm r}$=0.19 & $\sigma_{\rm r}$=0.21 & $\sigma_{\rm r}$=0.19     & ---  &   $\sigma_{\rm r}$=0.17 & $\sigma_{\rm r}$=0.33 & $\sigma_{\rm r}$=0.18  \\
\hline
$log(T_{\rm eff, MAX})$ & $r=$0.10  &   $r=$0.10 & $r=-$0.14 & $r=$0.27  & $r=$0.71  &   $r=$0.77 & $r=$0.90 & $r=$0.20  \\
    & $N$=8  & $N$=17 & $N$=8 & $N$=25 & $N$=8  & $N$=17 & $N$=8 & $N$=25    \\
       & $\sigma_{\rm r}$=0.44  &   $\sigma_{\rm r}$=0.25 & $\sigma_{\rm r}$=0.34 & $\sigma_{\rm r}$=0.20     & $\sigma_{\rm r}$=0.26  &   $\sigma_{\rm r}$=0.16 & $\sigma_{\rm r}$=0.14 & $\sigma_{\rm r}$=0.21  \\
\hline
$\Delta(V-I)$ & $r=$0.21  &   $r=$0.66 & $r=$0.64 & $r=$0.76     & $r=$0.49  &   $r=$0.57 & $r=$0.75 & $r=$0.10  \\
   & $N$=15   & $N$=38  & $N$=21 & $N$=59 & $N$=15   & $N$=38  & $N$=21 & $N$=59 \\
       & $\sigma_{\rm r}$=0.26  &   $\sigma_{\rm r}$=0.12 & $\sigma_{\rm r}$=0.14 & $\sigma_{\rm r}$=0.08     & $\sigma_{\rm r}$=0.23  &   $\sigma_{\rm r}$=0.14 & $\sigma_{\rm r}$=0.14 & $\sigma_{\rm r}$=0.14  \\
\hline
\hline                                                                          
\end{tabular}\\                       
}                                         
$^{*}$ GCs with double SGB excluded from the analysis\\ 

\caption{Spearman's rank correlation coefficients indicating the statistical dependence between $L1$ (columns 2-5), $L2$ (columns 6-9) and several parameters of the host GCs for G1, G2, G3, and G2+G3 GCs. The values of $\sigma_{\rm r}$, which provide an estimate of the robustness of $r$ measurements, and the numbers of analyzed GCs ($N$) are also listed.}
\end{table}

\begin{table}[!htp]
\center
\scriptsize {
\begin{tabular}{lcc}
\hline
\hline
  Author   & $age_{\rm G2}$ [Gyr] & $age_{\rm G3}$ [Gyr]  \\
\hline
           & all  &  \\
\hline
  Dotter et al.\,(2010, 2011) & 12.3$\pm$0.1 &  13.2$\pm$0.1 \\
  Mar{\'{\i}}n-Franch et al.\,(2009)     & 12.8$\pm$0.2 &  13.4$\pm$0.2 \\
   De Angeli et al.\,(2005)   & 12.3$\pm$0.3 &  13.2$\pm$0.3 \\
   VandenBerg et al.\,(2013) and Leaman et al.\,(2013) & 11.8$\pm$0.1 & 12.5$\pm$0.1 \\
\hline
           & metal-rich sample  &  \\
\hline
  Dotter et al.\,(2010, 2011)  & 12.1$\pm$0.2 &  12.9$\pm$0.3 \\
  Mar{\'{\i}}n-Franch et al.\,(2009)    & 12.7$\pm$0.3 &  13.3$\pm$0.1 \\
   De Angeli et al.\,(2005)    & 11.6$\pm$0.5 &  12.8$\pm$0.1 \\
   VandenBerg et al.\,(2013) and Leaman et al.\,(2013) & 11.4$\pm$0.2 & 12.3$\pm$0.5 \\
\hline
           & metal-intermediate sample  &  \\
\hline
  Dotter et al.\,(2010, 2011)  & 11.9$\pm$0.2 &  13.0$\pm$0.1 \\
  Mar{\'{\i}}n-Franch et al.\,(2009)    & 12.5$\pm$0.3 &  13.3$\pm$0.3 \\
  De Angeli et al.\,(2005)   & 11.7$\pm$0.5 &  13.0$\pm$0.5 \\
   VandenBerg et al.\,(2013) and Leaman et al.\,(2013) & 11.6$\pm$0.1 & 12.2$\pm$0.2 \\
\hline
           & metal-poor sample  &  \\
\hline
  Dotter et al.\,(2010, 2011)  & 13.1$\pm$0.1 &  13.3$\pm$0.1 \\
  Mar{\'{\i}}n-Franch et al.\,(2009)   & 13.0$\pm$0.3 &  13.4$\pm$0.3 \\
  De Angeli et al.\,(2005)   & 13.5$\pm$0.3 &  14.0$\pm$0.5 \\
  VandenBerg et al.\,(2013) and Leaman et al.\,(2013) & 12.4$\pm$0.1 & 12.7$\pm$0.1 \\
\hline
\hline
\end{tabular}
}
\caption{Average ages for G2 and G3 GCs.}
\label{tab:age}
\end{table}
\end{document}